\documentclass[twocolumn]{aastex61}
\usepackage{amsmath}

\newcommand{\be}{\begin{enumerate}}
\newcommand{\ee}{\end{enumerate}}

\shorttitle{}
\shortauthors{Gallegos-Garcia et al.}

\begin{document}

\title{Tidal Disruptions of Main Sequence Stars of Varying Mass and Age: \\ Inferences from the Composition of the Fallback Material}

\correspondingauthor{Monica Gallegos-Garcia}

\email{mpgalleg@ucsc.edu}

\author{Monica Gallegos-Garcia}
\affiliation{Department of Astronomy and Astrophysics, University of California, Santa Cruz, CA 95064, USA}

\author{Jamie Law-Smith}
\affiliation{Department of Astronomy and Astrophysics, University of California, Santa Cruz, CA 95064, USA}
\affiliation{Niels Bohr Institute, University of Copenhagen, Blegdamsvej 17, 2100 Copenhagen, Denmark}

\author{Enrico Ramirez-Ruiz}
\affiliation{Department of Astronomy and Astrophysics, University of California, Santa Cruz, CA 95064, USA}
\affiliation{Niels Bohr Institute, University of Copenhagen, Blegdamsvej 17, 2100 Copenhagen, Denmark}

\begin{abstract}
We use a simple framework to calculate the time evolution of the composition of the fallback material onto a supermassive black hole arising from the tidal disruption  of main sequence stars. We study stars with masses between 0.8 and 3.0 $M_\sun$, at evolutionary stages from zero-age main sequence to terminal-age main sequence, built using the Modules for Experiments in Stellar Astrophysics code.  We show that most stars develop enhancements in nitrogen ($^{14}$N) and depletions in carbon ($^{12}$C) and oxygen ($^{16}$O) over their lifetimes, and that these features are more pronounced for higher mass stars. We find that, in an accretion-powered tidal disruption flare, these features become prominent only after the time of peak of the fallback rate and appear at earlier times for stars of increasing mass. We postulate that no severe compositional changes resulting from the fallback material should be expected near peak for a wide range of stellar masses and, as such, are unable to  explain the extreme helium-to-hydrogen line ratios  observed in some TDEs. On the other hand, the resulting compositional changes could help explain the presence of nitrogen-rich features, which are currently only detected after peak. When combined with the shape of the light curve, the time evolution of the composition of the fallback material provides a clear method to help constrain the nature of the disrupted star. This will enable a better characterization of the event by helping  break the degeneracy between the mass of the star and the mass of the black hole when fitting tidal disruption light curves.  
\end{abstract}


\section{Introduction}\label{sec:intro}

Tidal disruption events (TDEs) offer a way to study both galactic supermassive black holes (SMBHs) and the dense stellar clusters that surround them. In these clusters, each star traces out a complicated orbit under the combined influence of the SMBH and all the other stars. The orbits slowly diffuse as a result of the cumulative effect of stellar encounters \citep{1999MNRAS.309..447M}.
There is a chance that one of these interactions will rapidly shift a star onto a nearly radial orbit, bringing it close to the SMBH. If a star wanders too close to the SMBH it can be violently ripped apart by the SMBH's tidal field \citep[e.g.,][]{1988Natur.333..523R}. As a result, for a full disruption, about half of the disrupted  material eventually falls back and accretes onto the SMBH. This accretion is expected to power a flare that contains vital information about the disruption and can be used to constrain the properties of the SMBH and the disrupted object \citep{1976MNRAS.176..633F}.

The disruption of stars by SMBHs has been linked  to tens  of flares in the cores of previously quiescent  galaxies \citep{2017ApJ...838..149A,2015JHEAp...7..148K}.
Transient  surveys  such as  the  Palomar  Transient  Factory  (PTF), the All-Sky Automated Survey for Supernovae (ASAS-SN) and the  Panoramic  Survey  Telescope  and  Rapid  Response  System  (Pan-STARRS)  are now finding  increasing numbers of these events, especially at early times
\citep{2014ApJ...793...38A, 2014MNRAS.445.3263H, 2012Natur.485..217G}.
By capturing the rise, peak, and decay of the flares, and with the addition of spectroscopic information, these events are starting to provide significant information about the underlying mechanisms \citep[e.g.,][]{2014ApJ...783...23G}.

Modeling TDEs properly requires a prediction of the rate of mass return to the SMBH after a disruption. While previous numerical results have provided reasonably precise models for the fallback resulting from the disruption of stars \citep[e.g.,][]{2013ApJ...767...25G}, 
they are incomplete in that they do not directly examine the predicted compositional changes.\footnote{Except for the specific case of a helium white dwarf with hydrogen envelope \citep{2017ApJ...841..132L}.} Additionally, many previous studies have focused on stars of a single structural profile, usually selected to match the Sun. However, typical stellar mass functions in TDE host galaxies predict that tidal disruptions should commonly involve evolved main sequence stars \citep{
2014ApJ...793...38A,
2016ApJ...818L..21F,
2017ApJ...835..176F,
2017ApJ...850...22L} whose internal structures are very diverse. 

Given that the accretion time is inferred to be significantly shorter than the period of the returning debris in most events, the fallback rate is expected to track the flare luminosity relatively closely \citep{1989ApJ...346L..13E, 2009MNRAS.400.2070S, 2009ApJ...697L..77R, 2014ApJ...783...23G}.
As the number of observed disruptions increases, and as the cadence and quality of data continues to improve, it has become increasingly important to improve models of the fallback material for disruptions of all kinds.

The presence or absence of particular emission line features in the spectra of TDEs might be used as a probe of the nature of the disrupted star
\citep{2017MNRAS.465L.114W, 2016ApJ...818L..32C, 2018MNRAS.473.1130B, 2017MNRAS.466.4904B, 2016MNRAS.462.3993B, 2016MNRAS.463.3813H, 2016MNRAS.455.2918H, 2016NatAs...1E..34L, 2015MNRAS.452...69M, 2015MNRAS.452.4297B, 2014MNRAS.445.3263H, 2014ApJ...793...38A, 2012ApJ...753...77C, 2012A&A...541A.106S, 2012Natur.485..217G}.
Motivated by this, in this paper, we expand upon work by \citet{2016MNRAS.458..127K}
to further characterize the rate of fallback and, in particular, the composition of the fallback debris. Our results predict what happens when stars of different masses and evolutionary states are tidally disrupted, and what composition a distant observer might be able to infer as the  signature of such events. 

In Section~\ref{sec:methods}, we briefly review the calculation of the mass accretion rate, $\dot{M}$, onto the SMBH, originally derived by \citet{2009MNRAS.392..332L}, and propose a simple generalization that allows $\dot{M}$ to be estimated from realistic stars. In Section~\ref{sec:results}, using this new framework, we present the accretion rate for stars ranging in mass from 0.8--3.0 $M_\sun$ and in evolutionary state from zero-age main sequence to terminal-age main sequence. In Section~\ref{sec:discussion}, we summarize our findings and discuss how our models can help inform the emission models of tidal disruption events by providing detailed predictions of the abundance of the radiating material.

\section{Methods}\label{sec:methods}
%----------------------------------------------------------------------------------------
\subsection{The Mass Accretion Rate} \label{sec:analyticdescription}
If a star with mass $M_{\star}$ and radius $R_{\star}$ is on a parabolic orbit around a SMBH of mass $M_{\rm bh}$ with pericenter distance, $r_{\rm p}$, less than the tidal radius, $r_{\rm t} = R_{\star}(M_{\rm bh}/M_{\star})^{1/3}=R_{\star}q^{-1/3}$, the star will be tidally disrupted. Here $q\equiv M_{\star}/ M_{\rm bh}$ is the mass ratio.

When a star is disrupted, the debris moves on approximately ballistic trajectories, with a spread in specific orbital energy that is roughly frozen at $r_{\rm t}$. This spread arises because at the time of disruption, the leading portions of the star are deeper in the potential of the SMBH than the trailing portions, which are farther away. The spread in specific energy of the debris, $E_{\rm t}$, can be approximated by taking the Taylor expansion of the SMBH's potential at the star's location: 
\begin{equation}
E_{\rm t}=GM_{\rm bh}R_{\star}/r_{\rm t}^2=q^{-1/3}E_{\star}, 
\end{equation}
where $E_{\star}=GM_{\star}/R_{\star}$ is the specific self-binding energy  of the star. 
Because most stars that are tidally disrupted in galactic nuclei approach the SMBH on nearly zero  energy orbits, $E_{\rm t}$ determines the fallback timescale for the most tightly bound debris
\begin{equation}
\begin{split}
t_{\rm t} & ={\frac{\pi}{M_\star}} \left( \frac{M_{\rm bh} R_\star^3 }{ 2G}\right)^{1/2} \\
& =0.1\ {\rm yr}\left( \frac{M_{\rm bh}}{ 10^6 M_\sun}\right)^{1/2}\left( \frac{M_\star } {M_\sun}\right)^{-1}\left( \frac{R_\star} {R_\sun}\right)^{3/2}.
\label{eq:time}
\end{split}
\end{equation}

In order to form an accretion flow, the bound stellar debris  must lose a significant amount of energy by viscous dissipation \citep{2015ApJ...809..166G,
2016MNRAS.461.3760H,
2016MNRAS.455.2253B,
2015ApJ...804...85S}. 
If the viscosity is large enough to allow accretion onto the SMBH on a timescale shorter than $t_{\rm t}$, the  luminosity of the flare is expected to follow the rate of mass fallback $\dot{M}=(dM/dE)(dE/dt)\propto t^{-5/3}$, where $dM/dE=M_\star/(2E_{\rm t})$ for a star on an initially parabolic orbit and $q\ll 1$ \citep{1988Natur.333..523R,
1989IAUS..136..543P}.
The $t^{-5/3}$ dependence of  TDE light curves relies on the assumption that the specific energy distribution of stellar debris 
$dE/dM$ is roughly flat with orbital specific energy, which  is only valid at late times \citep{2013ApJ...767...25G}. At early times, the assumption of constant $dM/dE$ is incorrect and depends sensitively on the  structure of the disrupted star \citep{2009MNRAS.392..332L, 2009ApJ...697L..77R} and the strength of the tidal interaction \citep{1993ApJ...410L..83L,
2009ApJ...705..844G, 2013ApJ...767...25G}.

\citet{2009MNRAS.392..332L} and \citet{2012PhRvD..86f4026K} moved beyond this simple description by constructing models that explicitly calculate the energy distribution of the disrupted stellar debris to $\mathcal{O}(q^{1/3})$ for stars described by a self-gravitating, spherically symmetric, polytropic fluid. By solving the Lane-Emden equation they determined the density profile of the star, which in turn allowed them to calculate $dM/dE$. In this paper we build on their work and show how their formalism can be easily extended to estimate the rate at which the debris falls back to pericenter and is subsequently accreted for tidally disrupted stars with realistic profiles. 

\begin{figure}[tbp]
\epsscale{1.31}
\plotone{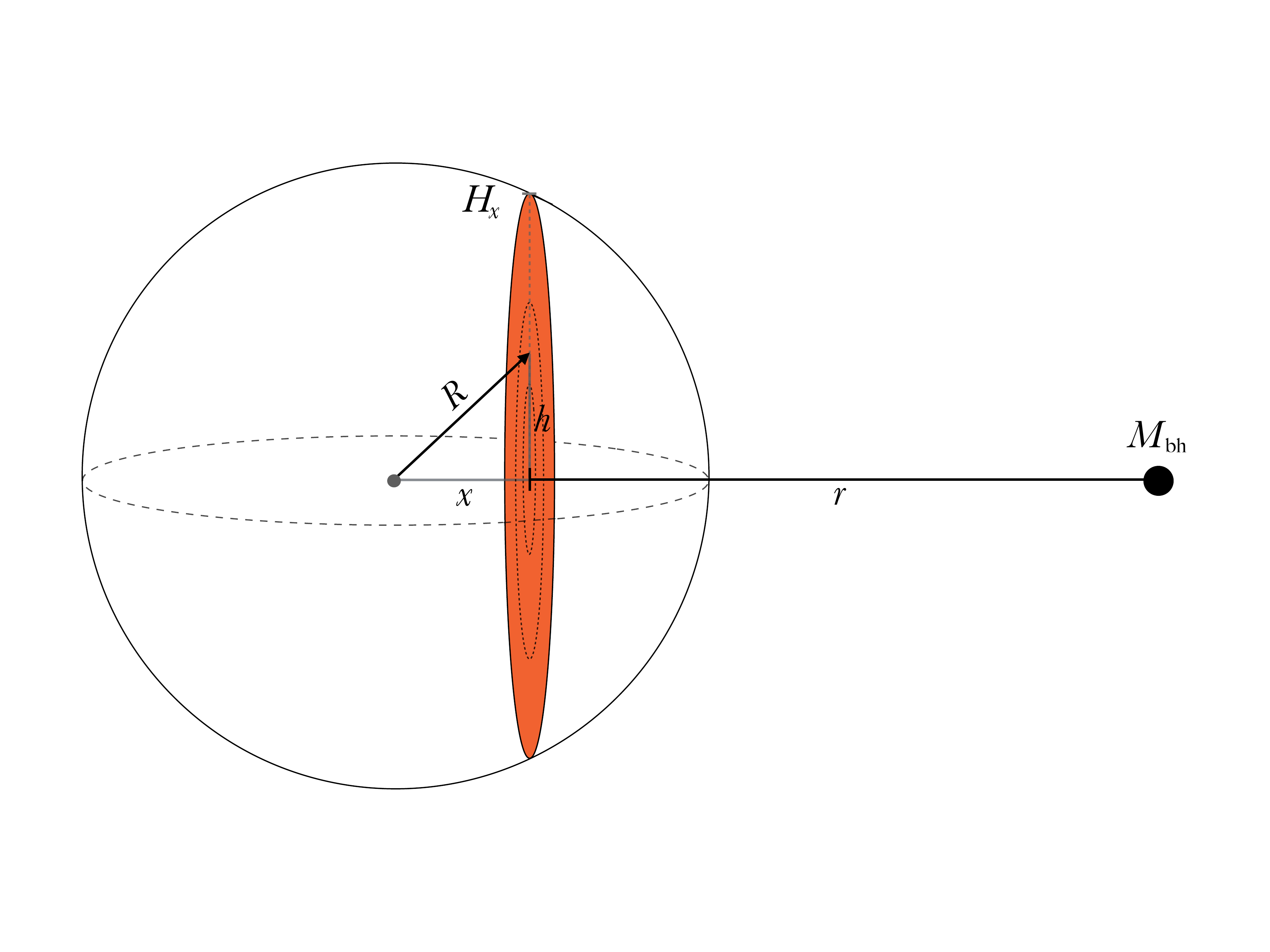}
\caption{The geometry of the disrupted star and how it can be used to calculate $dM/dE$. The orange slice represents an equal orbital binding energy surface, which can be approximated as an equal fallback time surface. Here $x$ is the distance from the center of the star along the star's orbital plane and $H_x$ is the maximum radius of the particular slice. When calculating the equal arrival time surfaces it is common to neglect any azimuthal or polar deviations. These can  be safely neglected given that  $(R_\star/r_{\rm t}) = q^{1/3} \ll 1 $.}
\label{fig:rr-hdiagram2}
\end{figure}

The geometrical setup envisioned here is shown in Figure~\ref{fig:rr-hdiagram2}. To calculate $\dot{M}$ we begin by using the standard assumption that the star {\it freezes in} at the moment of disruption at $r_{\rm t}$. The specific binding energy of a fluid element in this case depends on its position, and $dM/dE$ can be expressed in terms of the star's initial density profile $\rho_\star$. The mass of a slice of stellar debris $dM$, defined here as having the same orbital energy, is found by integrating 
\begin{equation}
\frac{dM}{dx}= \int_{0}^{H_x}  \rho_\star( h )  2\pi h \ dh,
\label{eq:integral}
\end{equation}
where $x$ is measured from the center of the star, $H_x$ is the radius of the slice at a given $x$, and $h$ is the rescaled height coordinate.  If the orbital period $t$ of a given slice is given in terms of its orbital binding energy $dE/dx$, then the rate $dM/dt$ at which mass falls back to pericenter can be calculated by numerically integrating  equation (\ref{eq:integral}). Using this  framework, we calculate the accretion rate history  for a large  number of realistic stars, whose density profiles we generate using the Modules for Experiments in Stellar Astrophysics (MESA) code. The reader is referred to Subsection \ref{sec:mesa} for a description  of our MESA setup.

The use of this analytic method allows for an extensive study of $\dot{M}$ arising from the disruption of different stars. While  this formalism leads to a large reduction in computational expense, it is nonetheless restricted as it relies on the assumption of a spherically symmetric star at the time of disruption. Contrary to what can be predicted by the simple analytical models used in this paper, the rate at which material falls back depends strongly on the strength of the encounter, which can be measured by the penetration factor $\beta \equiv r_{\rm t} / r_{\rm p}$. 

\begin{figure}[tbp]
\epsscale{1.30}
\plotone{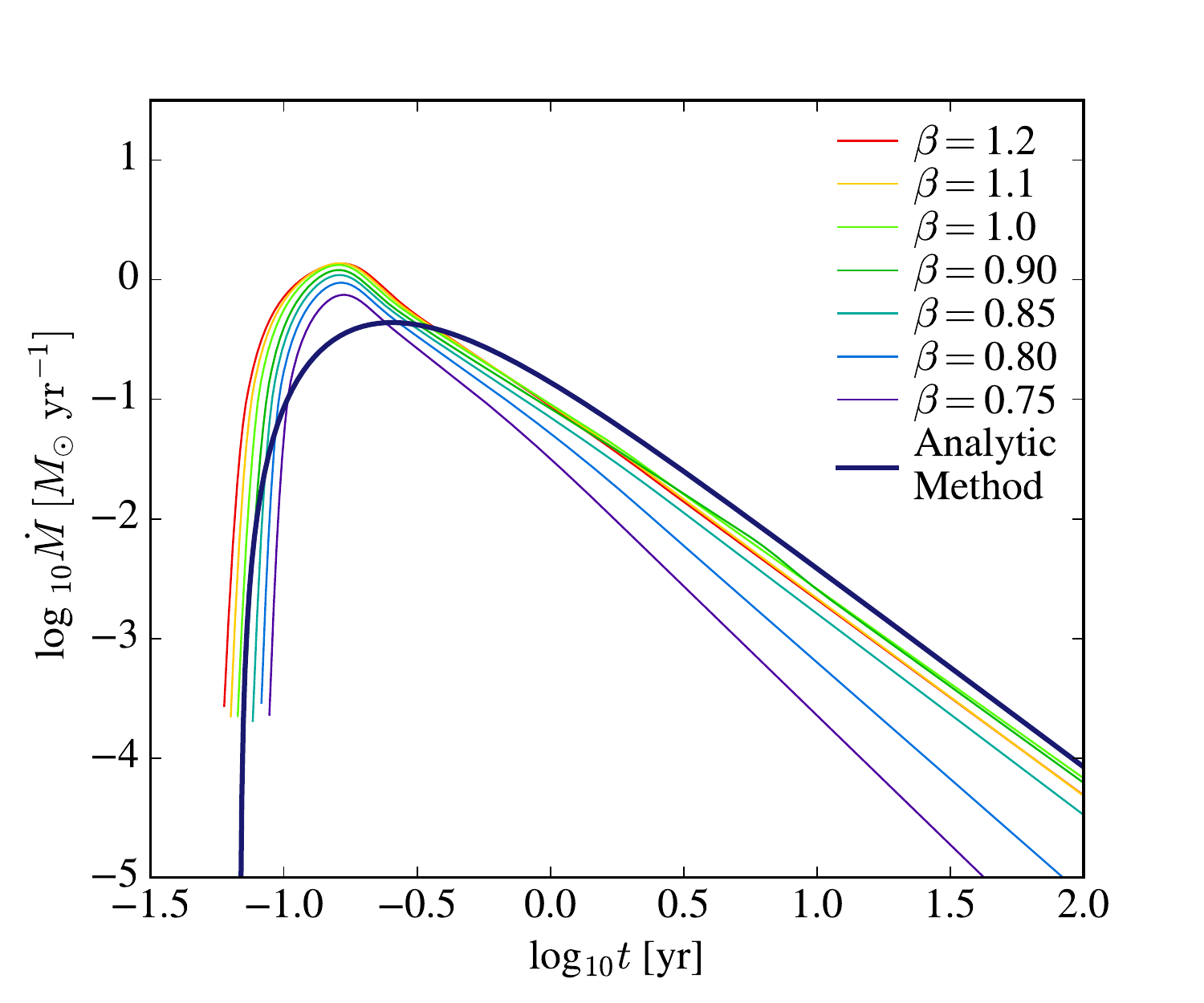}
\caption{The rate of fallback of stellar debris to pericenter as a function of time  from the disruption of a $1 M_\sun$ star calculated using the analytic framework used in this work ({\it thick dark blue} line), which assumes a full disruption, compared to those calculated by \citet{2013ApJ...767...25G} using hydrodynamical calculations for different $\beta$ values ({\it thin colored} lines).  Both calculations  use $M_\text{bh}=10^6M_\sun$ and a star that is constructed as  a self-gravitating, spherically symmetric, polytropic fluid with $\gamma = 5/3$. }
\label{fig:jamescomparison53}
\end{figure}

This is because varying $\beta$ changes the amount of mass lost by the star, which affects the rate at which the liberated stellar debris returns to pericenter \citep[e.g.,][]{2017ApJ...841..132L}. In Figure \ref{fig:jamescomparison53} we compare fallback curves calculated using the analytical model ({\it thick dark blue} line) to those calculated using simulations ({\it thin colored} lines). For the purpose of comparison, both models use a $1 M_{\sun}$ star with adiabatic index $\gamma = 5/3$ 
and a $10^{6} M_{\sun}$ SMBH. 
We find that the broad features of $\dot{M}$ are reasonably well captured by the simple model (the same holds true for stars constructed  with  $\gamma = 4/3$), as also argued by \citet{2009MNRAS.392..332L} and \citet{2012PhRvD..86f4026K}. This fact is extremely powerful in that it permits a reasonable characterization of TDE signatures without the need to run many computationally expensive simulations on the large set of stars we study here. 

What is more, for a fixed $\beta$, the time evolution of the forces applied is identical, regardless of the ratio of the star's mass to the mass of the SMBH. This is because the ratio of the time the star takes to cross pericenter to the star's own dynamical time depends only on $\beta$. Therefore, as long as $q\ll 1$, the tidal disruption problem is self-similar, and our results can be scaled to predict how the time (Equation~\ref{eq:time}) of peak accretion rate, $t_{\rm peak}$, and its corresponding  magnitude  $\dot{M}_{\rm peak}$ change with $M_{\rm bh}$, $M_{\star}$ and $R_{\star}$: 
\begin{equation}
\dot{M}_{\rm peak} \propto M_{\rm bh}^{-1/2}M_{\star}^{2} R_{\star}^{-3/2},
\label{eq:m-peak}
\end{equation}
and
\begin{equation} \label{eq:t-peak}
t_{\rm peak} \propto M_{\rm bh}^{1/2} M_{\star}^{-1} R_{\star}^{3/2}.
\end{equation}
This fact is extremely powerful in that it permits us to completely  characterize the properties of a disruption of a given star with one  calculation. An exception to these simple scalings is if the star penetrates deeply enough such that $r_{\rm p}$  is comparable to the Schwarzschild radius $r_{\rm g}$. In this case, general relativistic effects can alter the outcome, especially if the black hole is spinning \citep{1993ApJ...410L..83L,2012PhRvD..86f4026K}.

We remind the reader that the exact value of the time of peak accretion rate $t_{\rm peak}$ and its corresponding  magnitude  $\dot{M}_{\rm peak}$ are not precisely determined. Most of these differences arise from how the problem was originally formulated, in which the star's self-gravity is ignored, and only the spread in binding energy across the star at pericenter is assumed to be important to determining $\dot{M}$.
Our primary goal in this paper is to develop a robust formalism for calculating the rate of fallback and its associated chemical composition as well as conducting a preliminary survey of the key stellar evolution parameters associated with this problem. The formalism presented in this section is well suited to this goal.

%----------------------------------------------------------------------------------------
\subsection{Stellar Models}\label{sec:mesa}

We use the open source MESA code \citep{2011ApJS..192....3P} to calculate the structure and composition of the stars that will be disrupted. We generated 192 solar metallicity stellar profiles ranging in mass from 0.8--3.0 $M_\sun$ and evolutionary state from zero-age main sequence (ZAMS) to near terminal-age main sequence (TAMS). Profiles are spaced in intervals of 0.05 in central hydrogen fraction.

The MESA setup used here is described below.\footnote{Inlists are available upon request.} We begin with a pre-MS model, use the \texttt{mesa\_49} nuclear network with the \texttt{jina} rates preference, the \citet{2009ARA&A..47..481A}
abundances ($X$=0.7154, $Y$=0.2703, and $Z$=0.0142), and \texttt{mixinglengthalpha=2.0}. The final profile, which we call TAMS, is at a central hydrogen fraction of $10^{-3}$. Time steps are limited to a maximum change in central hydrogen fraction of $1\%$.

We consider the mass range of 0.8--3.0 $M_\sun$ as stars with masses below $0.8~M_\sun$ will not evolve appreciably over the age of the universe,
and stars with masses above $3~M_\sun$, with MS lifetimes  $<300$ Myr, are  unlikely to be disrupted (the relaxation time for most galactic nuclei is $\gg$ 300 Myr).

We do not consider evolved stars for two reasons. First, the contribution of evolved stars to the current and near-future tidal disruption population is expected to be modest \citep{2012ApJ...757..134M}. Second, studies of the tidal disruption of evolved stars such as \citet{2012ApJ...757..134M} have shown that even for large $\beta$, giant stars are effective at retaining envelope mass and effectively retaining their  cores (where the differences in composition arise from MS and post-MS evolution). In this paper we are interested in the evolved material in the inner-most layers of stars that can be reasonably revealed during a TDE and thus we do not focus on significantly evolved stars. 

\subsection{Salient Model Features}

\begin{figure}[tbp]
\epsscale{1.23}
\plotone{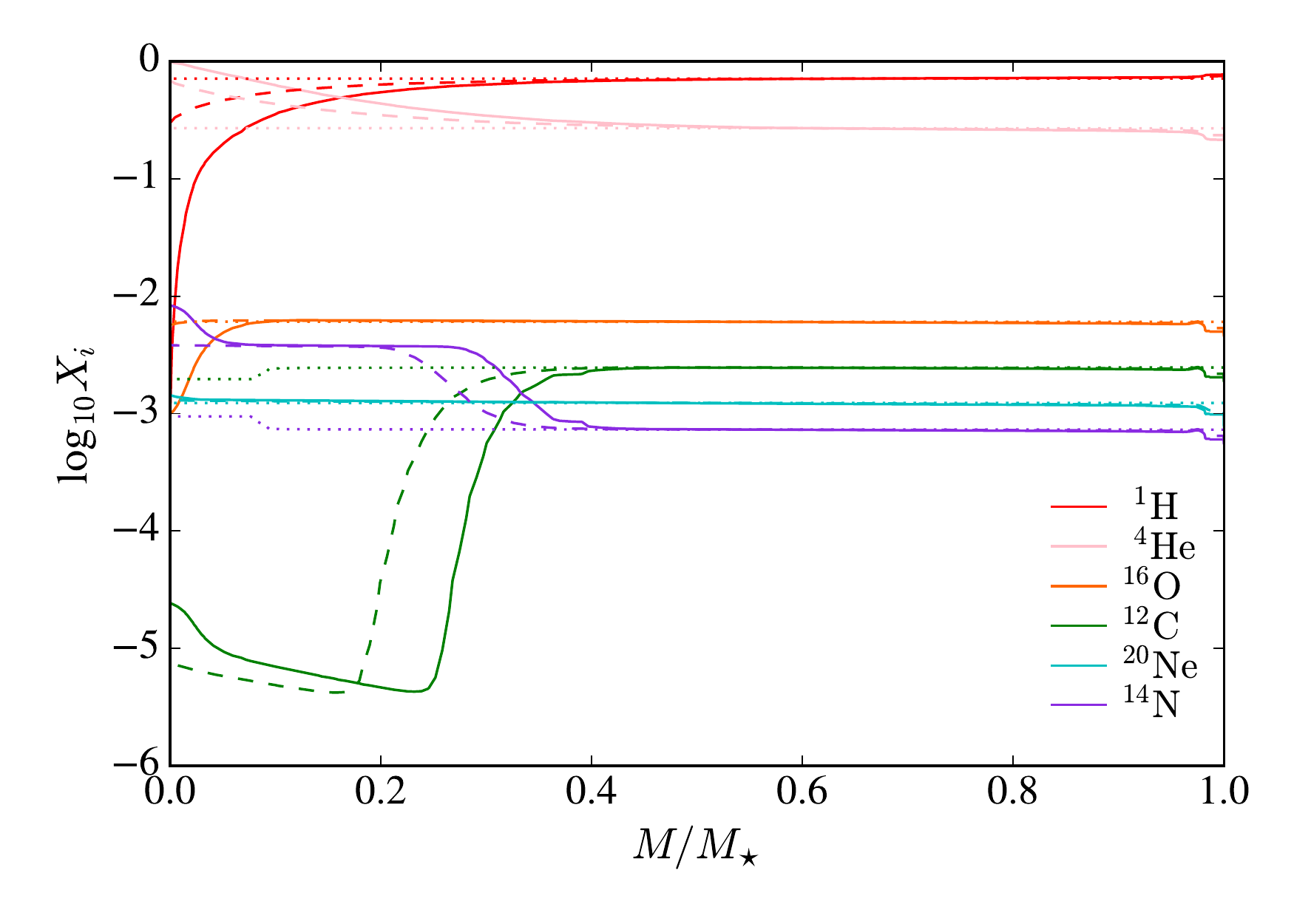}
\caption{Compositional abundance as a function of enclosed mass in a $1M_\sun$ star at three different evolutionary stages during its MS lifetime. In this paper, we characterize evolutionary stages by $f_{\rm H}$, the fraction of central hydrogen that has been burned. Here we show the stellar profiles for $f_{\rm H} = 0.0 = f_{\rm ZAMS}$ ({\it dotted}), $f_{\rm H}$ = 0.60 ({\it dashed}), and $ f_{\rm H} =  $ 0.99 ({\it solid}), respectively. A $1M_\sun$  star disrupted at later stages in its evolution should reveal abundance anomalies: an increase in nitrogen and depletion of oxygen, as previously argued by \citet{2016MNRAS.458..127K}.}
\label{fig:compositionallayering}
\end{figure}

Here we briefly discuss the stellar evolution features that are central to our study; these arise from changes in mass and evolutionary state along the MS. The two main burning processes in MS stars, the p-p chain and the CNO cycle, are highly sensitive to interior temperatures \citep{2012sse..book.....K} and contribute differently to stars of varying  mass. The p-p chain, which increases the abundance of $^{4}\rm{He}$ in stars, roughly dominates for masses $ \lesssim 1.5 \ M_\sun$. For masses  $\gtrsim 1.5 \ M_\sun$ the CNO cycle dominates. During the CNO cycle, fusing hydrogen to helium results in an increase (decrease) of $^{14}\rm{N}$ ($^{16}\rm{O}$) abundance, with $^{12}\rm{C}$  acting as a catalyst for the entire cycle. As argued by \citet{2016MNRAS.458..127K}, strong  compositional  variations are expected in the fallback material of MS stars. In this paper we trace the abundance variations of the following elements: $^{1}\rm{H}$,  $^{4}\rm{He}$, $^{16}\rm{O}$, $^{12}\rm{C}$, $^{20}\rm{Ne}$, and $^{14}\rm{N}$. These elements make up at least 99.6\% of each star's total mass. The $^{34}\rm{S}$ contribution and abundance ratio is very similar to that of $^{20}\rm{Ne}$ and is thus not explicitly shown in this paper. In what follows, we present abundances  relative to solar.

\begin{figure}[tbp]
\epsscale{1.2}
\plotone{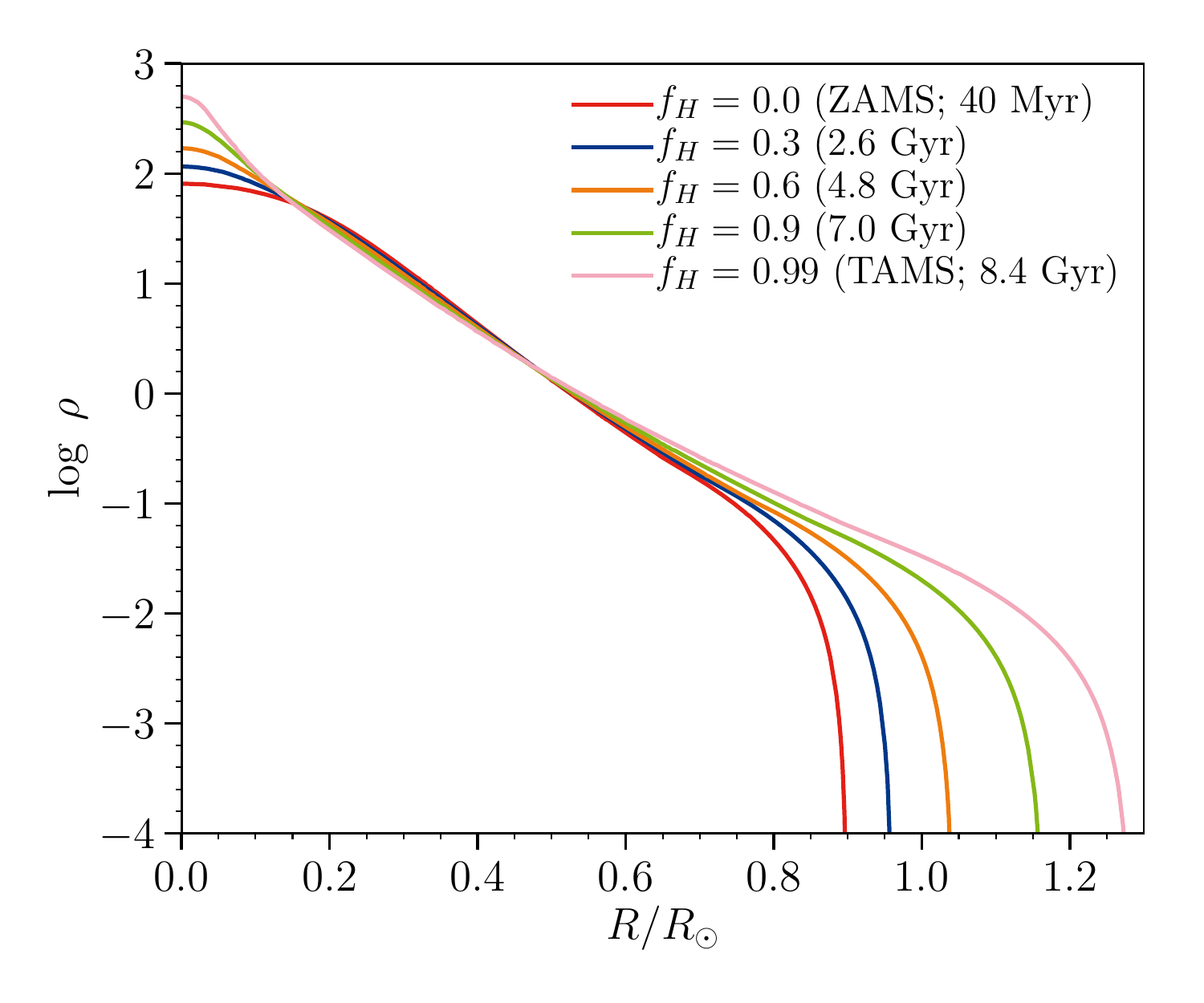}
\caption{Density profiles for a $1M_\sun$ star at different times along its  MS evolution. The {\it red} line corresponds to ZAMS with a central density of 81 $\rm g \ cm^{-3}$ and the {\it pink} line corresponds to a central hydrogen fraction of $10^{-3}$ with a central density of 500 $\rm g \ cm^{-3}$. These different density profiles result in different $r_{\rm t}$ and thus exhibit different vulnerability to disruption.}
\label{fig:mesaprofilesrhovsr}
\end{figure}

\begin{figure*}[tbp]
\epsscale{0.49}
\plotone{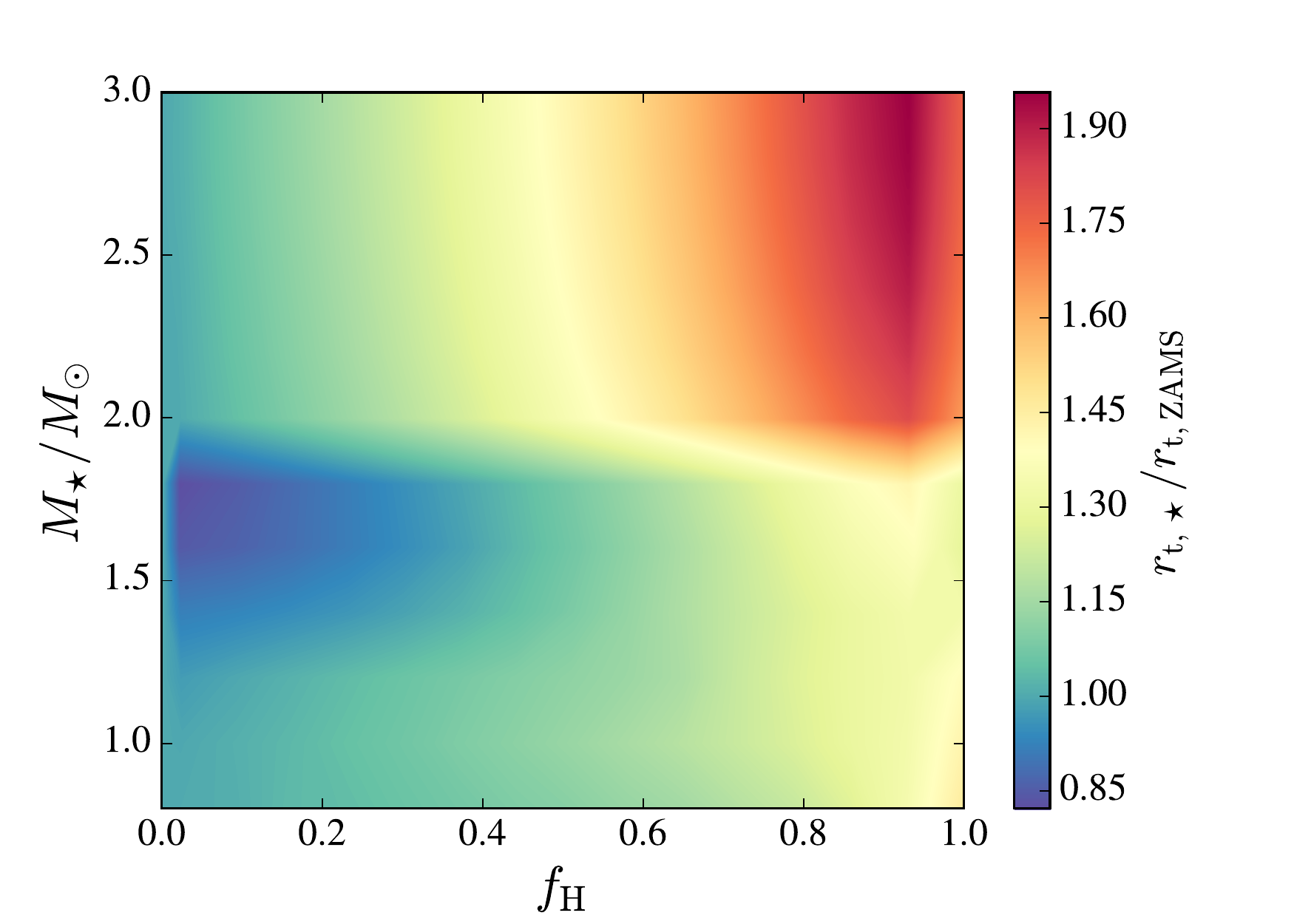}
\plotone{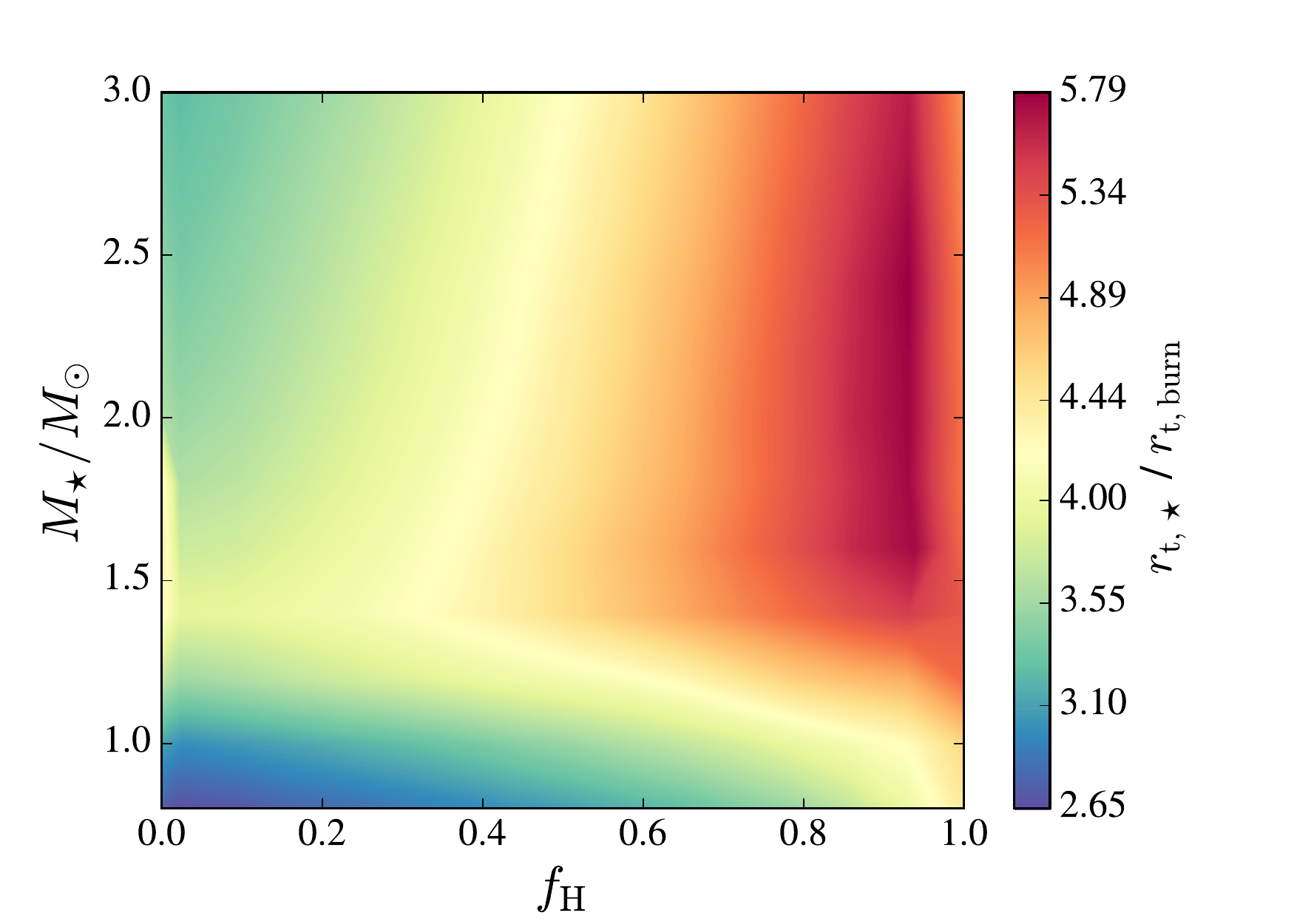}
\caption{In both panels, the color scale shows the tidal radius of the disrupted star. {\it Left panel}: Plotted are the ratio of the star's tidal radius to the tidal radius of that same  star at ZAMS ($f_{\rm H} = 0.0 = f_{\rm ZAMS}$). This shows that the star's vulnerability to disruption increases with age. This effect is stronger for more massive stars. {\it Right panel:} Plotted are the ratio of the tidal radius to $r_{\rm t, burn}$. Here $r_{\rm t, burn}$ is defined as the  tidal radius of the star's core undergoing active nuclear burning, where the specific power from nuclear reactions is greater than 1 erg g$^{-1}$s$^{-1}$. This shows that all of the stars in our study require deeper encounters to strip mass from their burning regions.}
\label{fig:tidalradius}
\end{figure*}

As an example, in Figure~\ref{fig:compositionallayering} we show the compositional variations along the MS for a $1 M_\sun$ star with solar abundance at ZAMS. The differently styled lines correspond to different stellar ages as defined by $f_{\rm H}$, the fraction of central hydrogen burned. A star will have $f_{\rm H}=0$ at ZAMS and  $f_{\rm H}=0.99$ near the end of its MS lifetime. At ZAMS the star has solar composition ({\it dotted} lines) and is roughly homogeneous. After  4.8 Gyr ({\it dashed} lines), when more than half of the central hydrogen has been processed ($f_{\rm H}=0.60$), the following abundance variations are seen: a significant increase of $^{14}\rm{N}$, a modest increase (decrease) of $^{4}\rm{He}$ ($^{1}\rm{H}$), a significant decrease of $^{12}\rm{C}$, and a roughly unchanged abundance of $^{20}\rm{Ne}$ and $^{16}\rm{O}$. At  TAMS ({\it solid} lines), where most of the central hydrogen has been processed ($f_{\rm H} = 0.99$), a  depletion in $^{16}\rm{O}$ abundance is also observed.
At this late stage, there is also a secondary increase in $^{14}\rm{N}$ in the core of the star.

In summary, we see that $^{1}\rm{H}$, $^{4}\rm{He}$ and $^{16}\rm{O}$ abundances evolve gradually, slowly extending to larger parts of the star and encompassing larger radii, while $^{12}\rm{C}$ and $^{14} \rm{N}$ abundances evolve rapidly across the burning region. 
All stars follow a similar trend. The most massive star in this study ($3M_{\sun}$) has, at TAMS, large compositional changes across roughly half of its mass (or about 20\% of its radius). As discussed by \citet{2016MNRAS.458..127K}, in the fallback material from a TDE we expect $^{12}\rm{C}$ and $^{14}\rm{N}$ abundance anomalies to be more noticeable and appear at earlier times than the other elemental anomalies. 

As a star evolves along the MS, its average density, $\bar{\rho}_{\star}$, decreases and its core density, $\rho_{\rm core}$, increases. This is illustrated  in Figure \ref{fig:mesaprofilesrhovsr}, where we show the evolution of the density profile for a $1M_{\sun}$ star with initial solar abundance from ZAMS to TAMS. Since the star's radius increases with age while its mass remains nearly constant, $\bar{\rho}_{\star}$ decreases with age. 
The effects of  $\bar{\rho}_{\star}$ on the star's vulnerability to tidal deformations can be readily seen by rewriting $r_{\rm t}$ as $r_{\rm t } \cong M_{\rm bh}^{1/3} \bar{\rho}_{\star}^{-1/3}$.  
This scaling implies that as the star evolves, it becomes progressively more vulnerable to tidal  deformations and mass loss. However, this scaling is unable to accurately capture the exact impact parameter required to fully disrupt a star. This is because as the star evolves a denser core, a surviving core is likely to persist for a disruption at $r_{\rm t}$ ($\beta = 1$), which is the penetration factor assumed for the analytical calculations. Nonetheless, we expect the time and magnitude of the peak accretion rate to be reasonably well captured by the simple formalism described here.

\section{The Disruption of Evolved MS Stars}\label{sec:results}
\subsection{Tidal Vulnerability}\label{sec:tidal-vulnerability}

Here we analyze how the tidal radius,  $r_{\rm t, \star}$, evolves with stellar mass and age along the MS for the stars in our study. The left panel of Figure \ref{fig:tidalradius} shows  $r_{\rm t, \star}$ normalized to the tidal radius of the same star at ZAMS, $r_{\rm t, ZAMS}$. We plot this ratio as a function of $f_{\rm H}$, the fraction of central hydrogen burned, and stellar mass $M_{\star}$. As expected, we find that the tidal radius increases with age and evolves more dramatically with $f_{\rm H}$ throughout the lifetime of more massive stars. For example, the tidal radius of a 3$M_\sun$ star increases by roughly a factor of two over its MS lifetime. As stars move along the MS, they become progressively more vulnerable to tidal dissipation and mass stripping. 

Next, we discuss how the vulnerability of regions with processed element abundances compares to that of the entire star. The right panel of Figure~\ref{fig:tidalradius} shows the ratio of $r_{\rm t, \star}$ to $r_{\rm t, burn}$, where $r_{\rm t, burn}$ is defined as the tidal radius of material within the regions of a star that exhibit active nuclear burning. This region of active nuclear burning is defined to be where the specific power from nuclear reactions is greater than 1 erg g$^{-1}$s$^{-1}$. This is a consistent way for defining the burning region  throughout all of the stellar profiles calculated here. As expected, this region is located  at small radii where the density is much higher than $\bar{\rho}_{\star}^{-1/3}$ and thus deeper penetrations are required in order to observe the evolved element abundances in the fallback material. Also, as this region is located within the innermost layers of the star, the processed elements will be revealed in the fallback material only at later times. 

\begin{figure}[tbp]
\epsscale{1.2}
\plotone{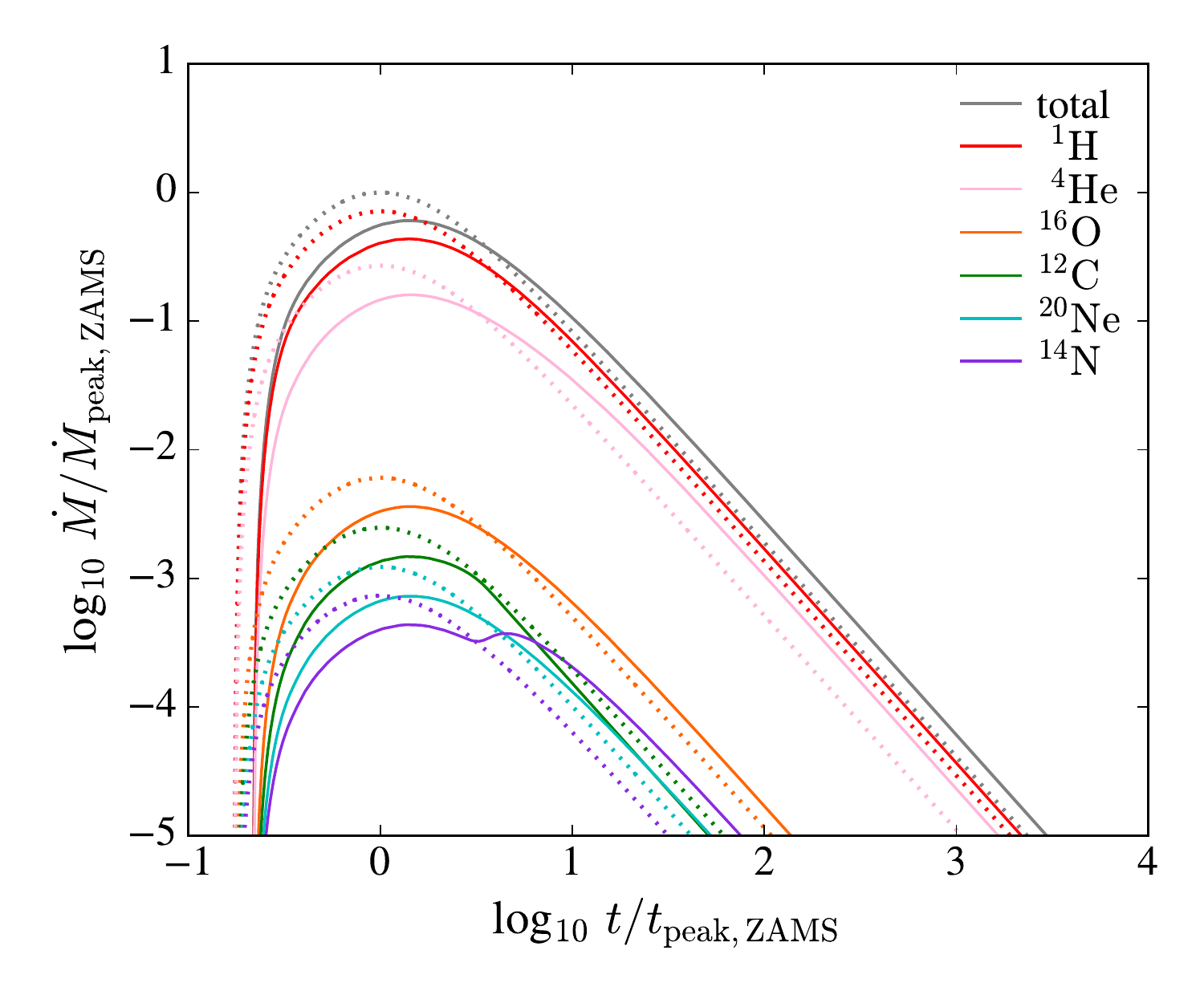}
\caption{Mass fallback rates for elements that make up 99.6\%  of the mass of a $1 M_\odot$ tidally disrupted  star at two different evolutionary stages. The star aged nearly 5 Gyr from the {\it dotted} lines ($f_{\rm H}=0.0=f_{\rm ZAMS} $) to the {\it solid} lines ($f_{\rm H} = $ 0.60). $\dot{M}$ for the total mass of the star is  shown by the {\it gray} curves. All curves are normalized to $\dot{M}_{\rm peak}$ and $t_{\rm peak}$ for the corresponding ZAMS star. The main changes in fallback rates as the star evolves along the MS are an increase in nitrogen and a decrease in carbon after $t_{\rm peak}$ due to CNO activity in the core.}
\label{fig:mdotnormzpeak}
\end{figure} 

\subsection{The Disruption of a Sun-like Star}\label{sec:Sun-like Star}
Figure~\ref{fig:mdotnormzpeak} shows the mass fallback rate arising from the full disruption of a $1 M_{\sun}$ star at two different evolutionary states: at ZAMS ({\it dotted} lines) and after 4.8 Gyr ({\it dashed} lines), when more than half of the central hydrogen has been processed ($f_{\rm H}=0.60$). These curves are normalized to the peak fallback rate and peak time of the corresponding ZAMS star: $\dot{M}_{\rm peak,  ZAMS}$ and $t_{\rm peak, ZAMS}$, respectively. 
The compositions of the stars before disruption are shown in Figure~\ref{fig:compositionallayering} as {\it dotted} (ZAMS) and {\it dashed} ($f_{\rm H} =  0.60$) lines. The disruption of the TAMS $1 M_{\sun}$ star, whose composition is shown by the {\it solid} lines in Figure~\ref{fig:compositionallayering}, is  expected to be similar in shape to the disruption of the $f_{\rm H}$ = 0.60 star, with an enhancement in $^{14} \rm N$ and depletion in $^{12} \rm C$. 

\begin{figure*}
\epsscale{0.9}
\plotone{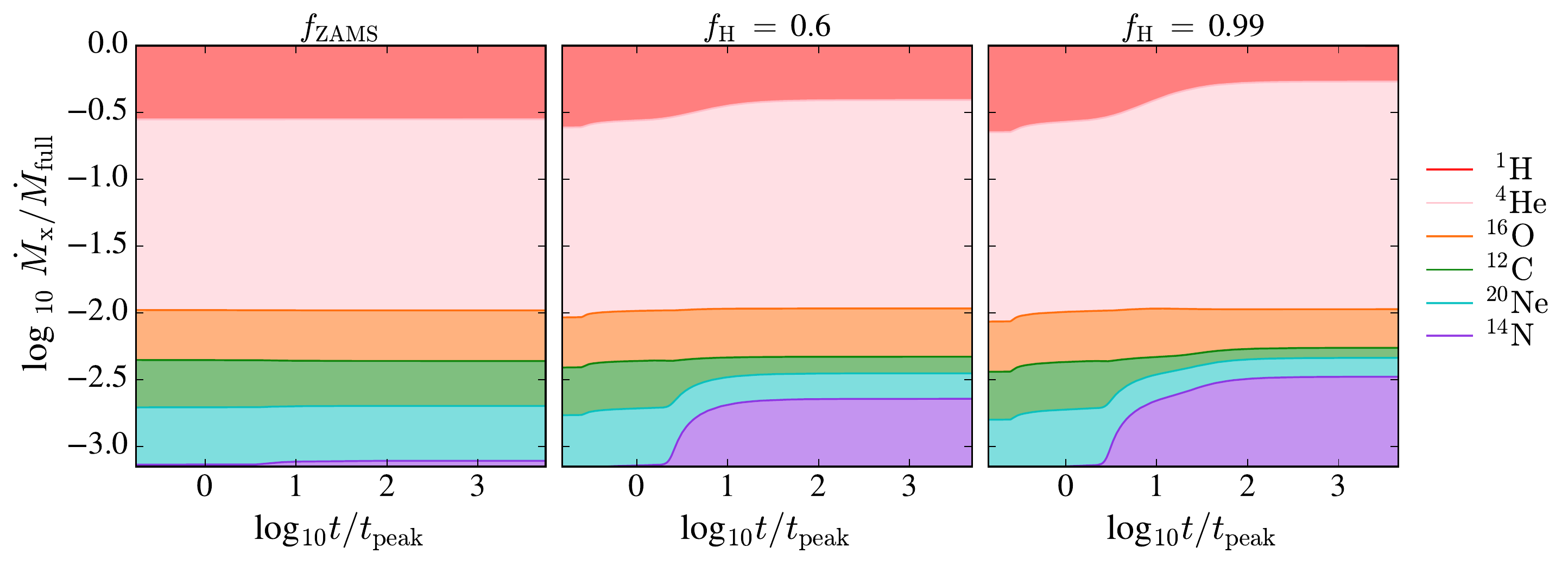}
\caption{The fallback rate for different elements, $ \dot{M}_{\rm X}$, following the disruption of a $ 1M_{\sun}$ star at three different evolutionary stages. 
The {\it left} and {\it center} panels correspond to the {\it dotted} and {\it solid} lines shown in Figure~\ref{fig:mdotnormzpeak}, respectively.
The {\it right} panel shows $\dot{M}$ for the same star but at  $f_{\rm H} = 0.99$, which corresponds to an age of 8.3 Gyr. Time is in units of $t_{\rm peak}$.  As the star ages we see an increase in nitrogen and a decrease in carbon abundance but only after $t_{\rm peak}$.}
\label{fig:mdotcontribution}
\end{figure*}

The smooth behavior of the fallback rates for all the plotted elements during the disruption of the ZAMS star ({\it dotted} lines in Figure~\ref{fig:compositionallayering}) is the result of the nearly homogeneous elemental composition within the star. The fallback rates for the $f_{\rm H}=0.60$ star ({\it dashed} lines in Figure~\ref{fig:compositionallayering}), on the other hand,  contain information about the varying nature of its elemental composition. In the fallback rates we can see an obvious increase in $^{14} \rm N$, decrease in $^{12} \rm C$, and a slight increase in $^{4} \rm He$, which is consistent with  the compositional structure of the star before disruption. These results are in agreement with \citet{2016MNRAS.458..127K}. We note that the fallback curves for the $f_{\rm H}=0.60$ star have no  abundance variations at $t \lesssim t_{\rm peak}$. These compositional anomalies might provide insight into the nature of the progenitor star near or after the most luminous time of the tidal disruption flare. 

In Figure~\ref{fig:mdotcontribution} we show the fractional contribution to the total fallback rate arising from each element during the disruption of a $1M_{\sun}$ star at three different evolutionary stages. From left to right, these panels correspond to the ZAMS ({\it dotted}), $f_{\rm H} = 0.60$ ({\it dashed}), and TAMS ({\it solid}) composition profiles in Figure~\ref{fig:compositionallayering}, respectively. In each panel we calculate the ratio of the fallback rate for each element, $ \dot{M}_{\rm X}$, to the total mass fallback rate, $ \dot{M}_{\rm full} $. 

For the disruption of a $1M_{\sun}$ star, it might be challenging to distinguish its evolutionary stage using spectral information if it is only obtained at $t \lesssim t_{\rm peak}$ (although the exact values of $\dot M_{\rm peak}$ and $t_{\rm peak}$ are expected to be  distinct; Figure~\ref{fig:LT}). This is, however, not the case after $t_{\rm peak}$. 

\begin{figure*}
\epsscale{0.8}
\plotone{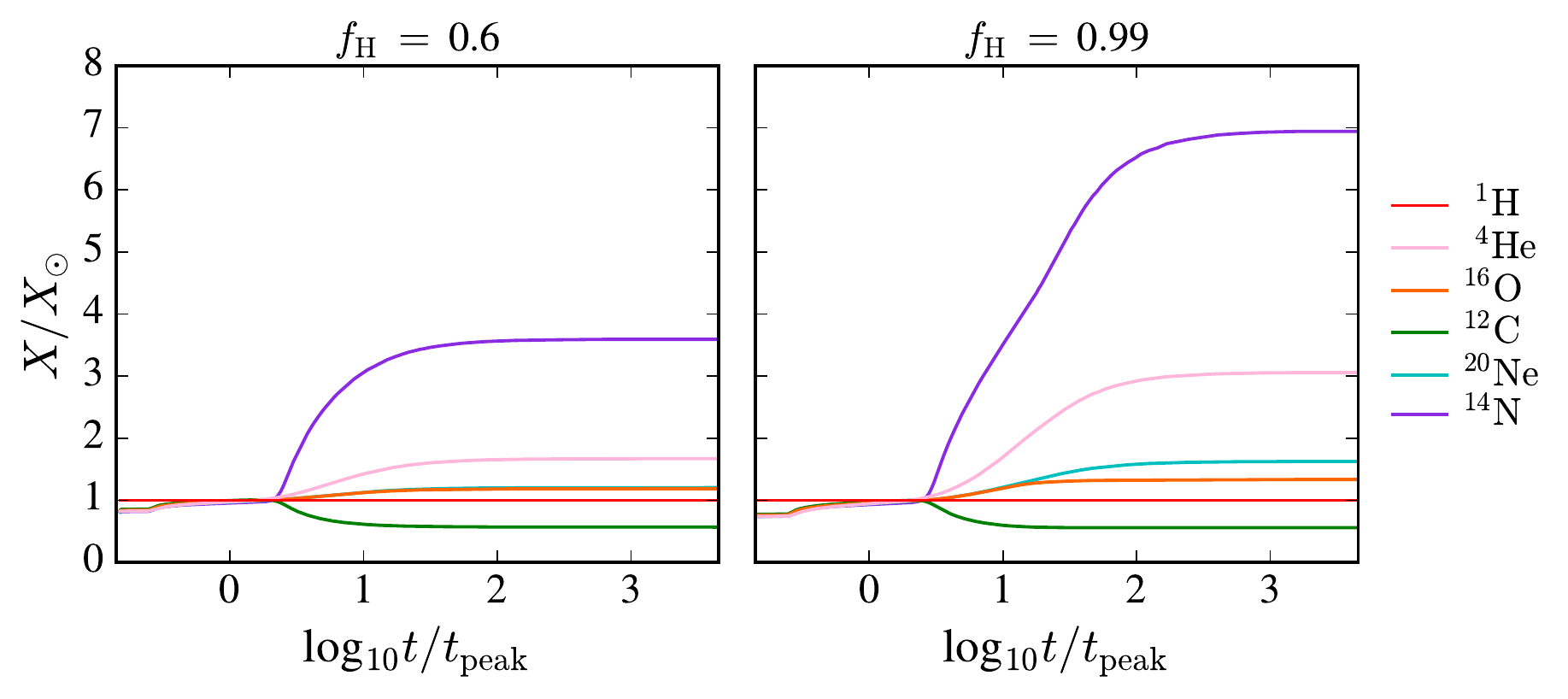}
\caption{Elemental abundance of the fallback material relative to solar following the disruption of a $1M_\sun$ at two different evolutionary stages: $f_{\rm H}=0.60$ ({\it left} panel)  and TAMS ({\it right} panel). A rapid evolution of $^{14} \rm N$ and $^{12} \rm C$ abundance relative to the other elements is clearly seen. The solar ratios clearly illustrate  the significance of the variations in the  abundances of $^{16} \rm O$, $^4 \rm He$, and $^{20} \rm Ne$.}
\label{fig:1Msun_xx}
\end{figure*}

Figure~\ref{fig:1Msun_xx} shows  the abundance of the fallback material  relative to solar following the disruption of a $1M_\sun$ at two different evolutionary stages: $f_{\rm H}=0.60$ ({\it left} panel)  and TAMS ({\it right} panel). Elemental abundances  relative to solar  are  calculated  here using 
\begin{equation}
\frac{ {X} }{ X_{\sun} } = 
\frac{ \dot{M}_{\rm X} / \dot{M}_{\rm H} }{  M_{\rm X} / M_{\rm H, \sun } },
\label{eq:solar}
\end{equation}
where $\dot{M}_{\rm X}$ is the fallback rate for a selected element, $\dot{M}_{\rm H} $ is the fallback rate of $^{1} \rm H$, and $M_{\rm X}/ M_{\rm H, \sun }$ is the abundance mass ratio relative to solar of element $\rm X$. 
The disruptions of a $f_{\rm H} = 0.60$ and a TAMS star each show a significant increase in $^{14} \rm N$ and $^4 \rm He$ after $t_{\rm peak}$. As expected, these features are more prominent for the TAMS star. Near $t = 10 t_{\rm peak}$, Figure \ref{fig:1Msun_xx} shows steeper abundance gradients in the {\it right} panel compared to the {\it left} in all  elements  except $^{12} \rm C$. We note that these values are relative to $ ^{1} \rm H$. This is important in the case of $^{16}\rm{O}$ and $^{20}\rm Ne$ where we see an increase in their abundance. 
This is because while $ ^{1} \rm H$ is depleted at every evolutionary stage, $^{16}\rm{O}$ and $^{20}\rm Ne$ abundance remain relatively constant for a star of this mass, which results in higher solar ratios. However, this behavior is also altered by the mass of the star as we discuss in the following  section.

\subsection{Disruption of MS stars}
\label{sec: evolved ms star disruption}
For reasons discussed previously, it seems likely that the evolutionary state of a star might be revealed by charting the compositional evolution of the fallback material, which might be inferred from particular features in the spectra of the resulting luminous flare. The association of a significant fraction of TDEs with post-starburst galaxies \citep{2014ApJ...793...38A,2016ApJ...818L..21F, 2017ApJ...835..176F, 2017ApJ...850...22L} has suggested the likely presence of evolved stars in the nuclei of TDE hosts, or at least a subset thereof.
Much of our effort in this section will thus be dedicated to determining the state of the fallback material after the tidal disruption of stars of  a wide range of ages and masses.

\begin{figure*}
\epsscale{0.9}
\plotone{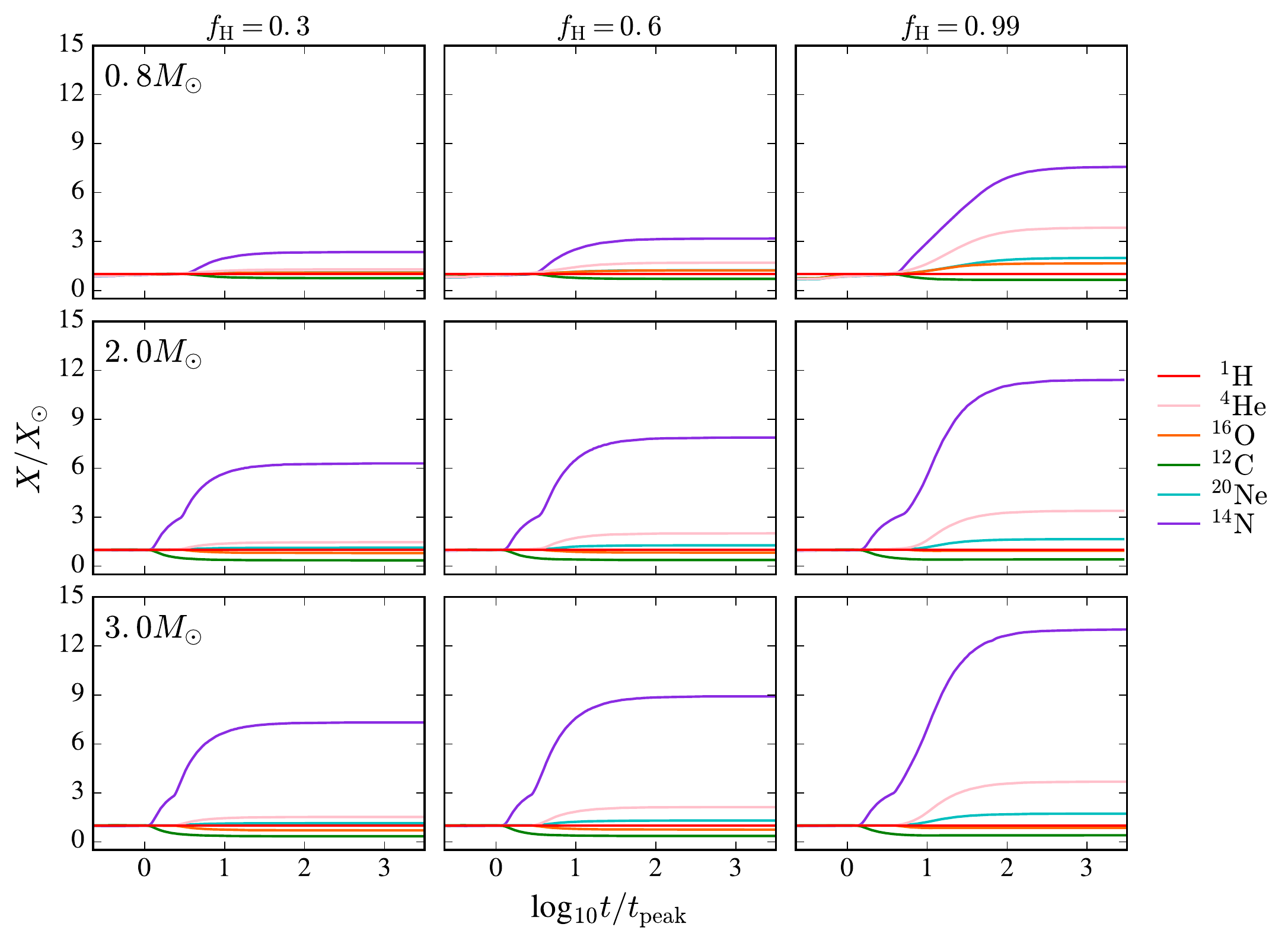}
\caption{The relative abundance of stellar debris as a function of fallback time arising from the disruption  of $0.8 M_{\sun}$ ({\it top} row), $2.0 M_{\sun}$ ({\it middle} row) and $3.0 M_{\sun}$ ({\it bottom} row) stars at three different evolutionary stages ($f_{\rm H}=0.3,0.6$ and 0.99). The change in abundance relative to solar is observed to increase with mass and age but only after $t_{\rm peak}$. These  anomalies appear at earlier times for higher mass stars.}
\label{fig:mdot_xx_3by3}
\end{figure*}

\begin{figure*}[tbp]
\epsscale{0.9}
\plotone{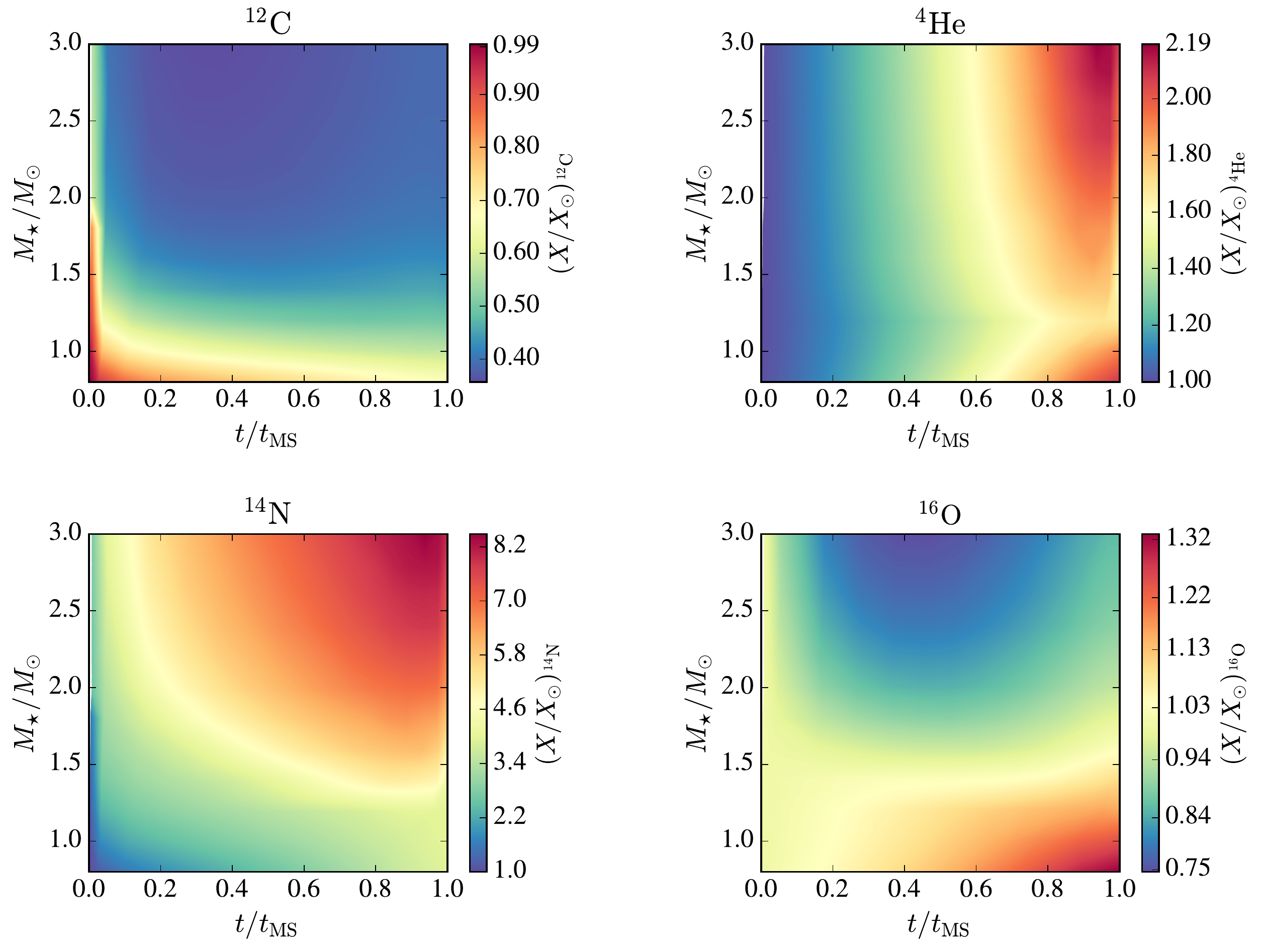}
\caption{Elemental abundances relative to solar at the time the mass fallback rate has reached one tenth of its peak value, $t_{0.1}>t_{\rm peak}$, for all of the stellar masses and ages in our sample. Elements of interest are $^{12}$C, $^{4}$He, $^{14}$N and $^{16}$O. Values are shown as a function of the star's fractional main sequence lifetime and stellar mass. We find carbon abundances to be more indicative of stellar mass for $M_\star \lesssim 1.5M_\sun$, while helium abundances are correlated with stellar age for all masses. $(X/X_{\sun})_{^{14} \rm N} \gtrsim 5.0$ occurs only for masses greater than $1.5M_\sun$ and develops early in the star's evolution. We also find oxygen abundances to be primarily stellar mass dependent.}
\label{fig:.10Mpeakabundance}
\end{figure*}

\begin{figure*}
\epsscale{0.9}
\plotone{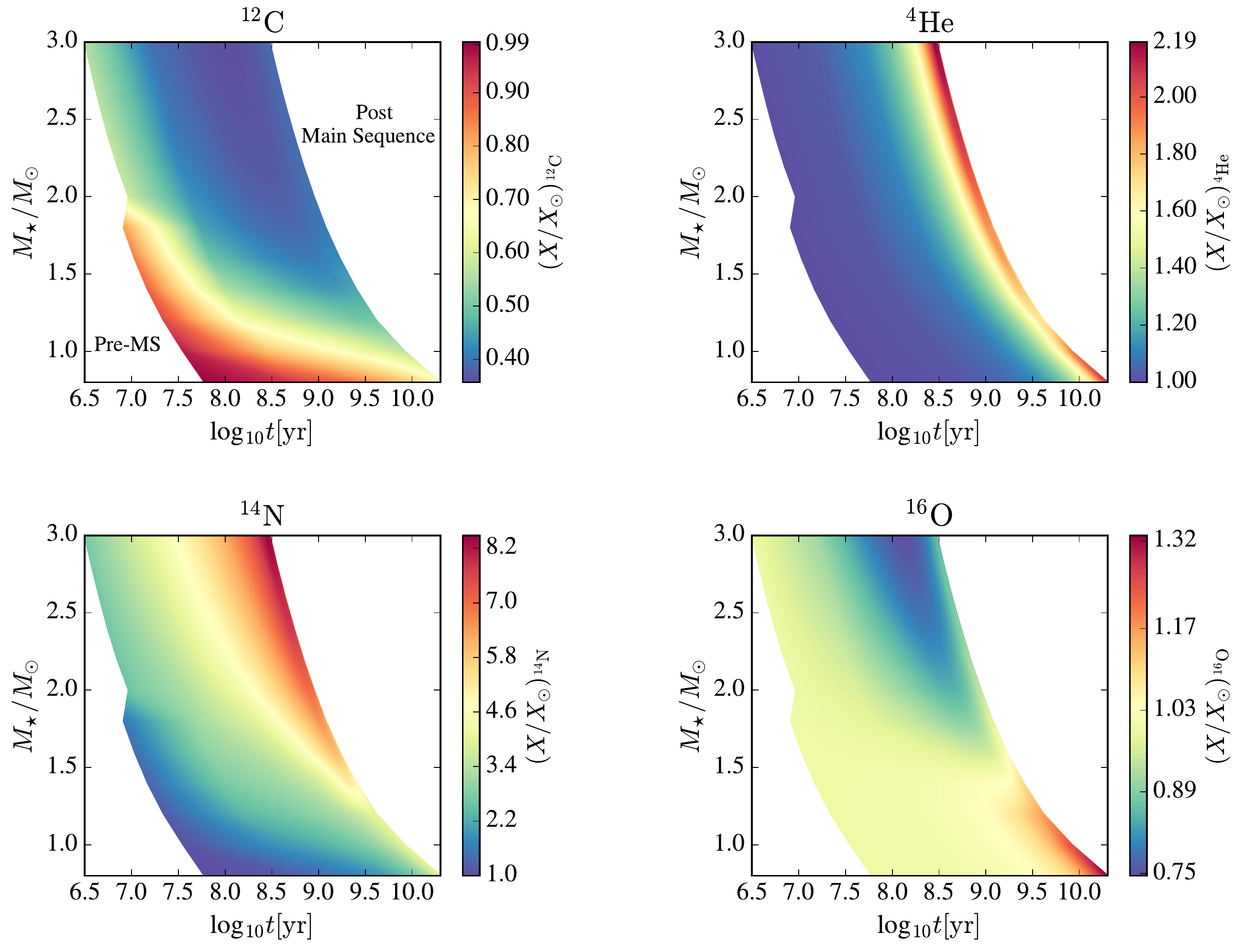}
\caption{Same as Figure \ref{fig:.10Mpeakabundance} but presented with stellar age in years ($x$-axis). The white regions correspond to pre-MS ({\it left}) or post-MS ({\it right}).}
\label{fig:0.1Mpeakabundance_realtime}
\end{figure*}

In Figure \ref{fig:mdot_xx_3by3} we show the relative abundances of the fallback material for three representative  MS star disruptions. The {\it first row} of panels shows the abundance of the fallback material for a $0.8 M_\sun$ star tidally disrupted at three different evolutionary stages: $f_{\rm H} = 0.3$, $f_{\rm H} = 0.6$, and $f_{\rm H} = 0.99$. The abundances shown are  similar to those shown in Figure \ref{fig:1Msun_xx} for a $1 M_\sun$ star. At these low masses, we expect the abundance anomalies  to be present in the fallback material at a few times $t_{\rm peak}$. 

The {\it second} row of panels in Figure \ref{fig:mdot_xx_3by3} shows the relative abundances of the fallback material for  a disrupted $2 M_{\sun}$ star. The abundance patterns are broadly similar to those seen for the $0.8M_{\sun}$ and $1 M_{\sun}$ stellar disruptions. However, there are three main differences. First, in contrast to the observed  increase of $^{16}\rm{O}$ seen  in the $0.8 M_{\sun}$ and $1.0 M_{\sun}$ disruptions, a significant decrease in $^{16}\rm{O}$ abundance is observed. This is an indication of the increased CNO activity in the $2 M_{\sun}$ star. Second, two distinct bumps are seen in the evolution of the $^{14} \rm N$ abundance, contrary to its steady increase in the smaller mass disruptions. The first increase in $^{14} \rm N$ abundance (and the corresponding $^{12} \rm C$ depletion) is due to the local maximum of CNO burning that is located at roughly 20\% of the star's radius. There is also significant CNO and p-p chain activity in the star's core, which is revealed at later times in the fallback material, and leads to the relatively delayed increase in $^{4} \rm He$ and $^{20}\rm{Ne}$, the corresponding decrease of $^{16}\rm{O}$, and a secondary increase in $^{14} \rm N$. Third, abundance variations are observed significantly closer to $t_{\rm peak}$ in the $2M_{\odot}$ disruptions than in the $0.8M_{\odot}$ disruptions. This is a result of the more extended burning region within the star, whose material is revealed at earlier times following the disruption. 

In the {\it bottom} row of panels in Figure \ref{fig:mdot_xx_3by3}  we show the composition of the fallback material following the disruption of a $3 M_{\odot}$ star. The abundance variations in these fallback curves closely resemble those for the $2 M_{\odot}$ star, but with larger variations appearing at earlier times. The abundance variations presented in Figure \ref{fig:mdot_xx_3by3} for the few representative stars accurately describe the overall trends in our sample. These trends are illustrated in Figure~\ref{fig:.10Mpeakabundance}, in which various elemental abundances are shown at the time that the mass fallback rate has reached one tenth of its peak value, $t_{0.1}>t_{\rm peak}$.

The fallback abundances at $t_{0.1}$ are plotted in Figure~\ref{fig:.10Mpeakabundance} as a function of the star's fractional main sequence lifetime, $t/t_{\rm MS}$, and stellar mass. In Figure~\ref{fig:0.1Mpeakabundance_realtime} we show the same abundance values as in Figure~\ref{fig:.10Mpeakabundance} but presented with the evolutionary age of the star in years. Some key points should be emphasized. We find carbon decrements to be indicative of stellar mass, while helium enhancements are indicative of age. $(X/X_{\sun})_{^{14} \rm N} \gtrsim 5.0$ occurs only for masses greater than $1.5M_\sun$ and develops early in the star's evolution. This is due to the enhanced CNO activity inside the more massive stars in our sample. We also find oxygen abundances to be primarily dependent on stellar mass. 

\begin{figure*}
\epsscale{0.4}
\plotone{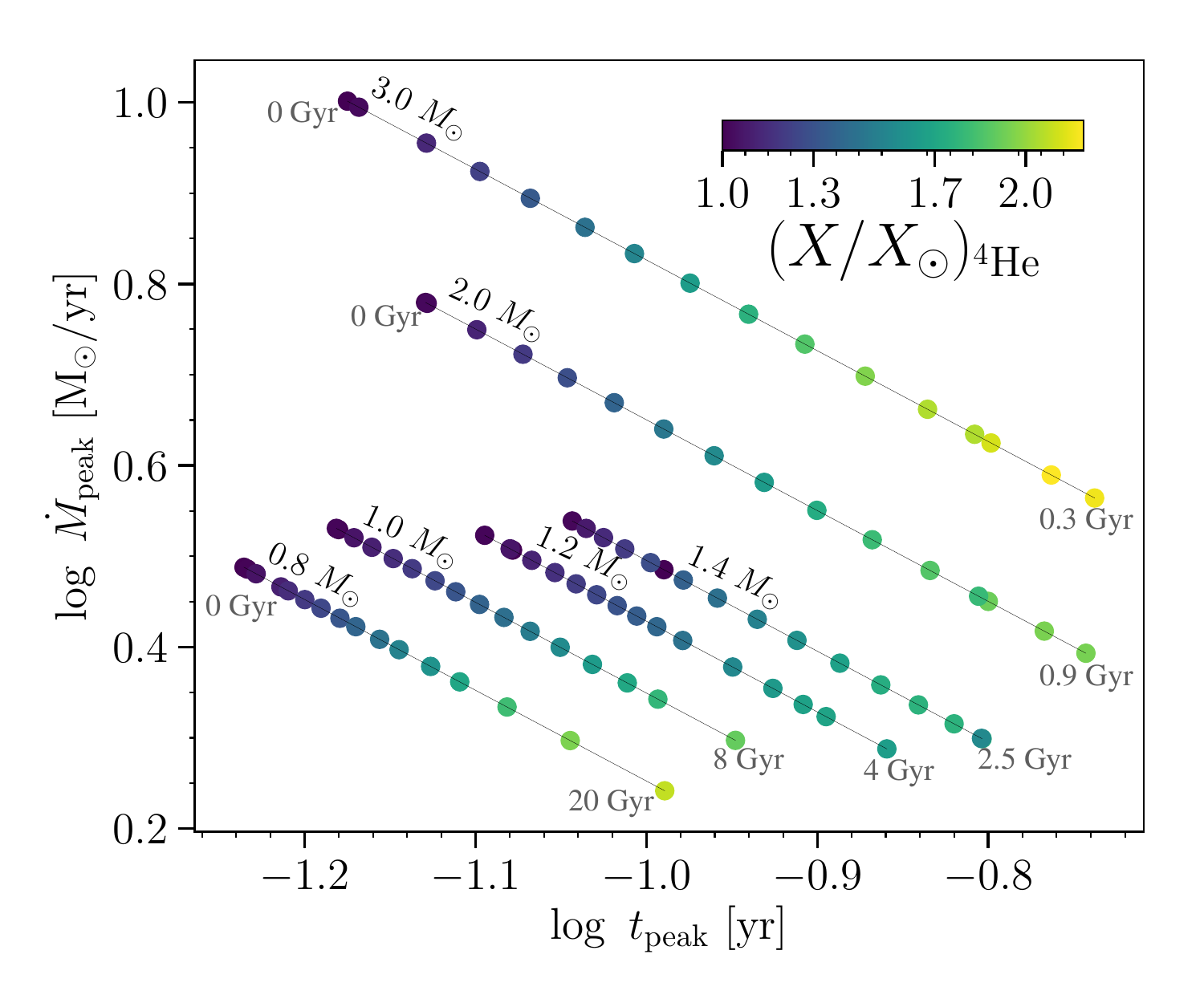}
\plotone{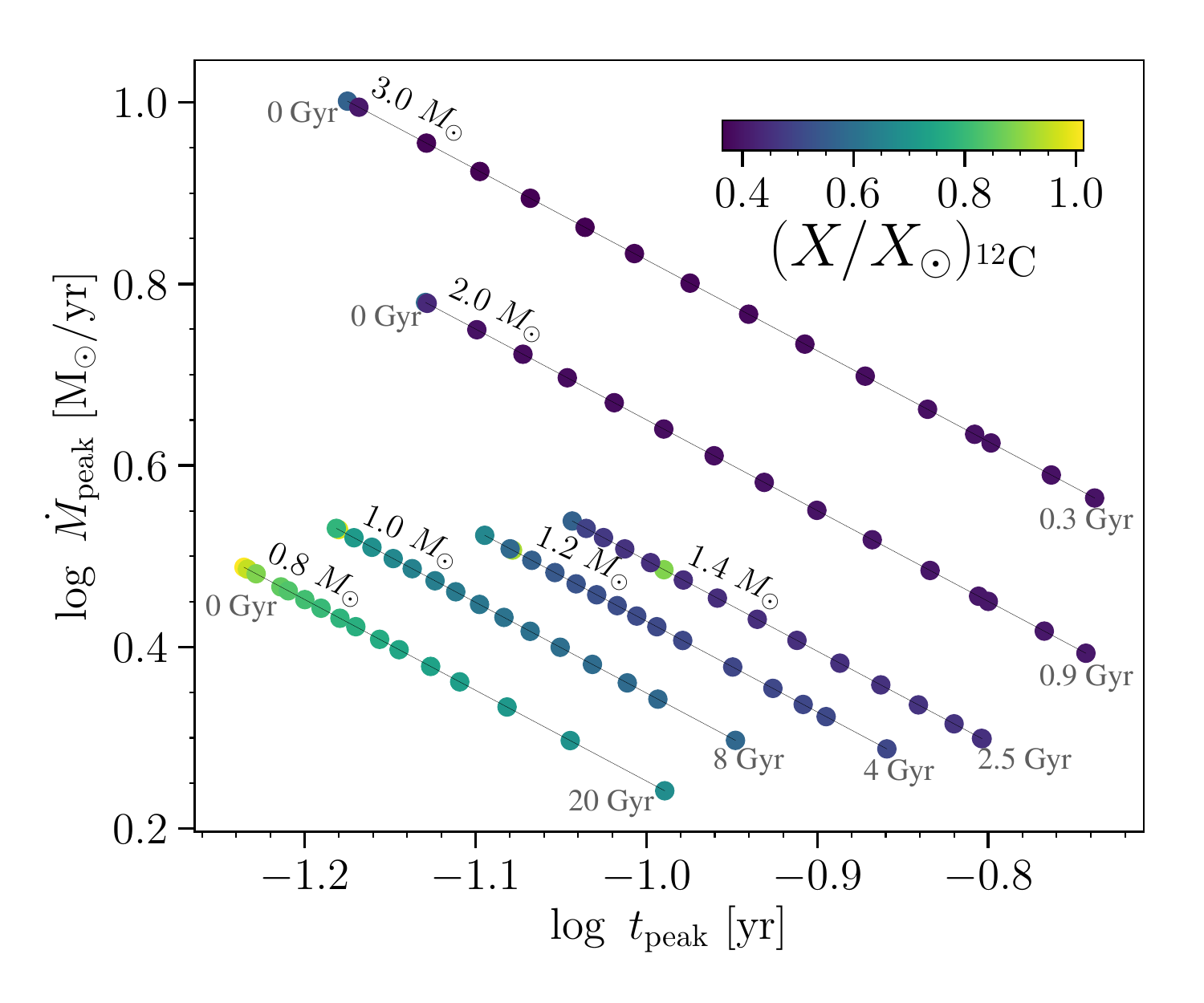}
\plotone{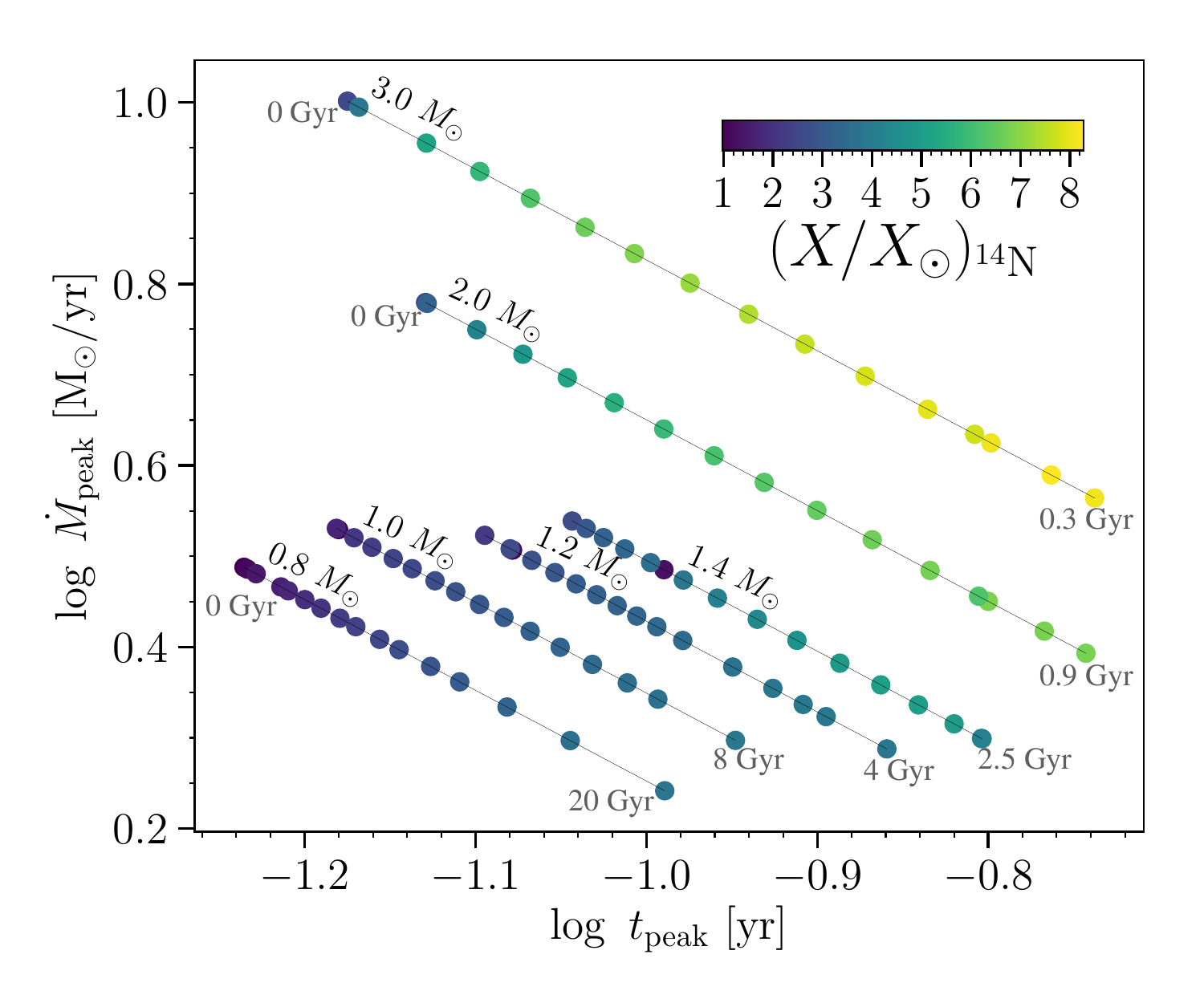}
\plotone{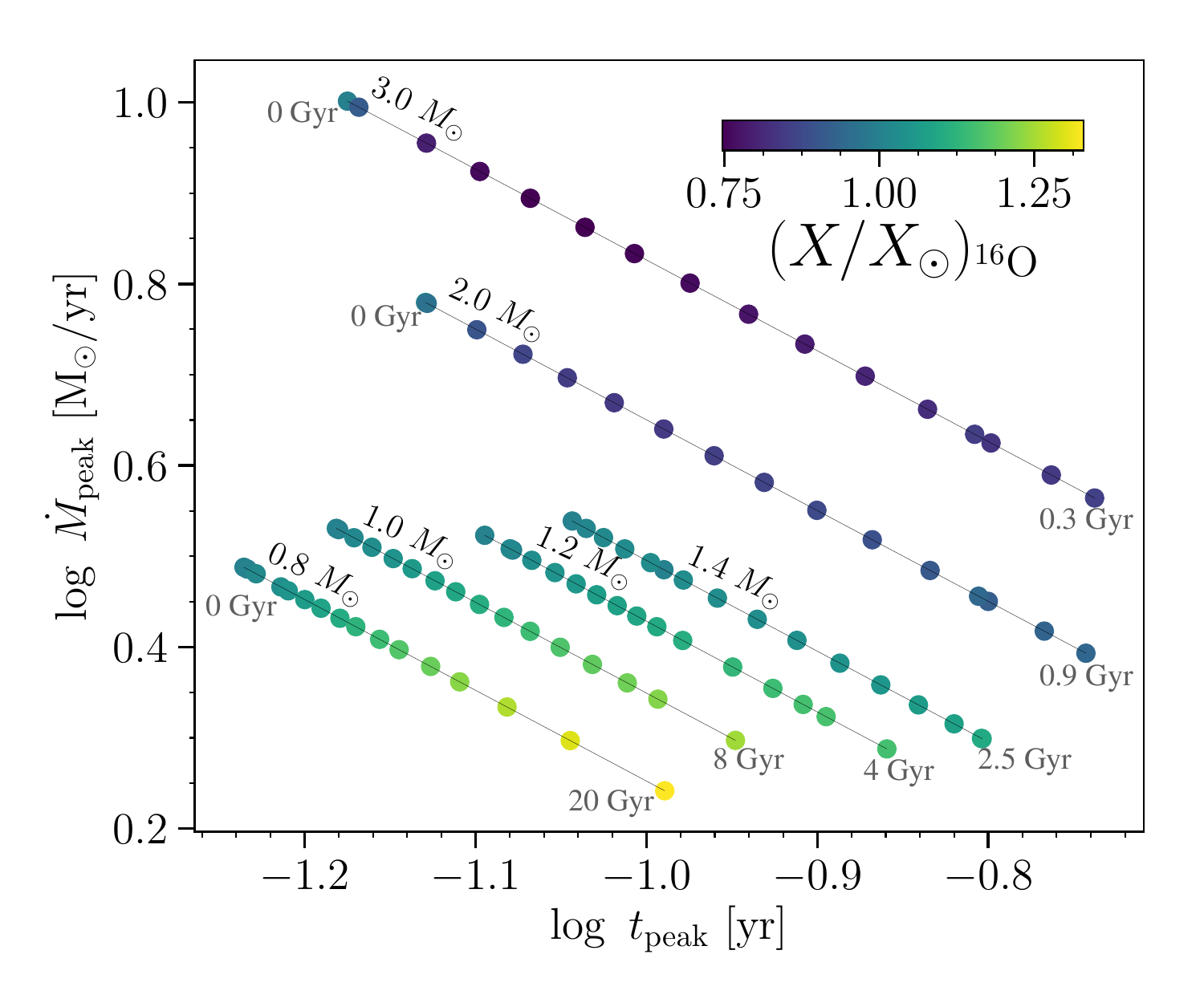}
\caption{Fallback abundance at $t_{0.1}$ of $^{4} \rm He$, $^{12} \rm C$, $^{14} \rm N$, and $^{16}\rm{O}$ (clockwise from top left) for the disruption (by a $M_{\rm bh}=10^6M_\odot$ SMBH) of 0.8$M_\sun$, 1$M_\sun$, 1.2$M_\sun$, 1.4$M_\sun$, 2.0$M_\sun$, and 3.0$M_\sun$ stars along their MS evolution. Abundances are at $t_{0.1}$, but points are placed at $\dot M_{\rm peak}$ and $t_{\rm peak}$ for the disruption of each star. Abundances are quoted relative to solar. Points are roughly equally spaced in time for each mass, with the top-left-most point being ZAMS and the bottom-right-most point being TAMS. (This is not strictly true for the ZAMS point of the 1$M_\sun$, 1.2$M_\sun$, and 1.4$M_\sun$ stars as their radius slightly decreases at the very beginning of their MESA evolution, but all other points for these stars proceed left to right with age as the star subsequently evolves.)
}
\label{fig:LT}
\end{figure*}

The processes discussed here suggest that TDEs may have a more complex spectrum and time-structure than simple models suggest. The effects are especially interesting when the accretion rate is high, as this gives rise to high luminosities, and thus can more readily offer clues to the nature of the disrupted star.  The specific values of $\dot{M}_{\rm peak}$ and ${t}_{\rm peak}$ can further aid in distinguishing the properties of the progenitor star before disruption. This is illustrated in Figure~\ref{fig:LT} where we show abundances of carbon, helium, nitrogen and oxygen (relative to solar) in the fallback debris as a function of $\dot M_{\rm peak}$ and $t_{\rm peak}$. Each panel in Figure~\ref{fig:LT} corresponds to a different element, the different lines correspond to different stars in our study ($0.8M_{\sun}$, $1.0M_{\sun}$, $1.2M_{\sun}$, $1.4M_{\sun}$, $2.0M_{\sun}$, and $3.0M_{\sun}$), the points are different stages in the stars' evolution on the MS (roughly equally spaced in time), and the color of the points is the abundance of the fallback debris at the time that $\dot M$ falls to one tenth of its peak value, $t_{0.1}$. We used the fitting formulas presented in \citet{2013ApJ...767...25G}, which give $\dot M_{\rm peak}$ and $t_{\rm peak}$ given $\beta$, $\gamma$, $M_\star$, and $R_\star$. We used $\gamma = 4/3$ and its corresponding penetration factor for full disruption  ($\beta = 1.85$) given by \citet{2013ApJ...767...25G}. The values of $M_\star$ and $R_\star$ were taken from the MESA profiles and we have assumed $M_{\rm bh}=10^6M_\odot$ (the reader is referred to equations~\ref{eq:m-peak} and \ref{eq:t-peak} for the scalings of  $\dot{M}_{\rm peak}$ and $t_{\rm peak}$ with $M_{\rm bh}$, respectively). The abundance values are the same as in Figure \ref{fig:.10Mpeakabundance}. 

The variation in elemental abundances is accompanied by a wide range in $\dot M_{\rm peak}$ and a moderate range in $t_{\rm peak}$; a combination of these different pieces of information can help characterize the progenitor stars of TDEs. For example, the disruption of a $3M_{\sun}$ star has similar $t_{\rm peak}$ values to that of a $2 M_{\sun}$ star. While their $^{12} \rm C$ and $^{16}\rm{O}$ abundances are very similar, the $3M_{\sun}$ star's disruption results in a higher abundance in $^{14} \rm N$ and $^{4} \rm He$ at every stage in its evolution, along with a higher $\dot M_{\rm peak}$. In the lower mass stars ($0.8$--$1.4 M_{\sun}$) there are many degeneracies in $\dot M_{\rm peak}$ and $t_{\rm peak}$ values. Here, the $^{14} \rm N$, $^{16}\rm{O}$, and $^{4} \rm He$ abundances are similar (over the age of the universe) but the $^{12} \rm C$ abundances vary at the early stages in these stars' MS evolution. Compositional information, combined with reprocessing and radiative transfer calculations \citep[e.g.,][]{2016ApJ...827....3R}, can thus be used to discern the stellar mass and age of the disrupted star. 

\section{Discussion}\label{sec:discussion}

\subsection{Summary of Key Results}
Motivated by the work of \citet{2016MNRAS.458..127K}, we have modeled the tidal disruption of MS stars of varying mass and age. We adopted the analytic formalism originally presented in \citet{2009MNRAS.392..332L} to study, for the first time, the time evolution of the composition of the fallback debris onto the SMBH. We compared the analytic method  to hydrodynamic simulations in Figure~\ref{fig:jamescomparison53} and found, similarly to \citet{2009MNRAS.392..332L} and \citet{2012PhRvD..86f4026K}, that the broad features of the fallback curves are reasonably well captured by it.\footnote{This work should, however, be taken only as a guide for the expected compositional trends in the fallback material, as hydrodynamical simulations are needed to accurately predict the evolution and characteristics of the flares.} We quantify the variations in composition arising from the disruption of 12 different stars with masses of $0.8$--$3.0M_{\sun}$ at 16 different evolutionary stages along the MS. The main results of our study are the following.  
\begin{enumerate}

\item We predict an increase in nitrogen and depletion in carbon abundance in the fallback debris with MS evolution for all stars in our sample \citep[in agreement with][]{2016MNRAS.458..127K}. We find a decrease in oxygen with MS evolution for $M_\star \gtrsim 1.5 M_\sun$, and an increase for $M_\star < 1.5 M_\sun$.

\item For all of the TDEs modeled in this study, we find that the time during the fallback rate curve when anomalous abundance features are present, $t_{\rm burn}$, is {\it after} the time of time of peak fallback rate $t_{\rm peak}$. 
 
\item Abundance variations are more significant and $t_{\rm burn} / t_{\rm peak}$ is smaller for stars of larger mass.

\item Some key variations in the compositional evolution  are highlighted, along with the types of observation that would help to discriminate between different stellar disruptions. In particular, we find carbon and oxygen abundances to strongly dependent on stellar mass for $M_\star \lesssim 2M_\sun$, while helium abundances are found to be correlated with stellar age for all masses. $(X/X_{\sun})_{^{14} \rm N} \gtrsim 5.0$ occurs only for masses greater than $1.5M_\sun$ and is observed  early in the star's evolution. 

\item Studying the compositional variation in the fallback debris provides a clear method for inferring the properties of the progenitor star before disruption. 
\end{enumerate}

\subsection{Implications for Observations and Models}
It is evident from the results described above that the evolution of the interior structure of stars during their MS lifetimes is very rich. Even in the simplest case of a Sun-like star, complex behavior with multiple abundance transitions in the fallback material may be observed. The resulting TDE spectra are expected to depend fairly strongly on the abundance properties of the fallback material \citep{2016ApJ...827....3R}. This implies that if one can be very specific about the times at which we expect to see such transitions in the observed emission, one can better constrain the properties of the disrupted star. 

\begin{figure}
\epsscale{1.1}
\plotone{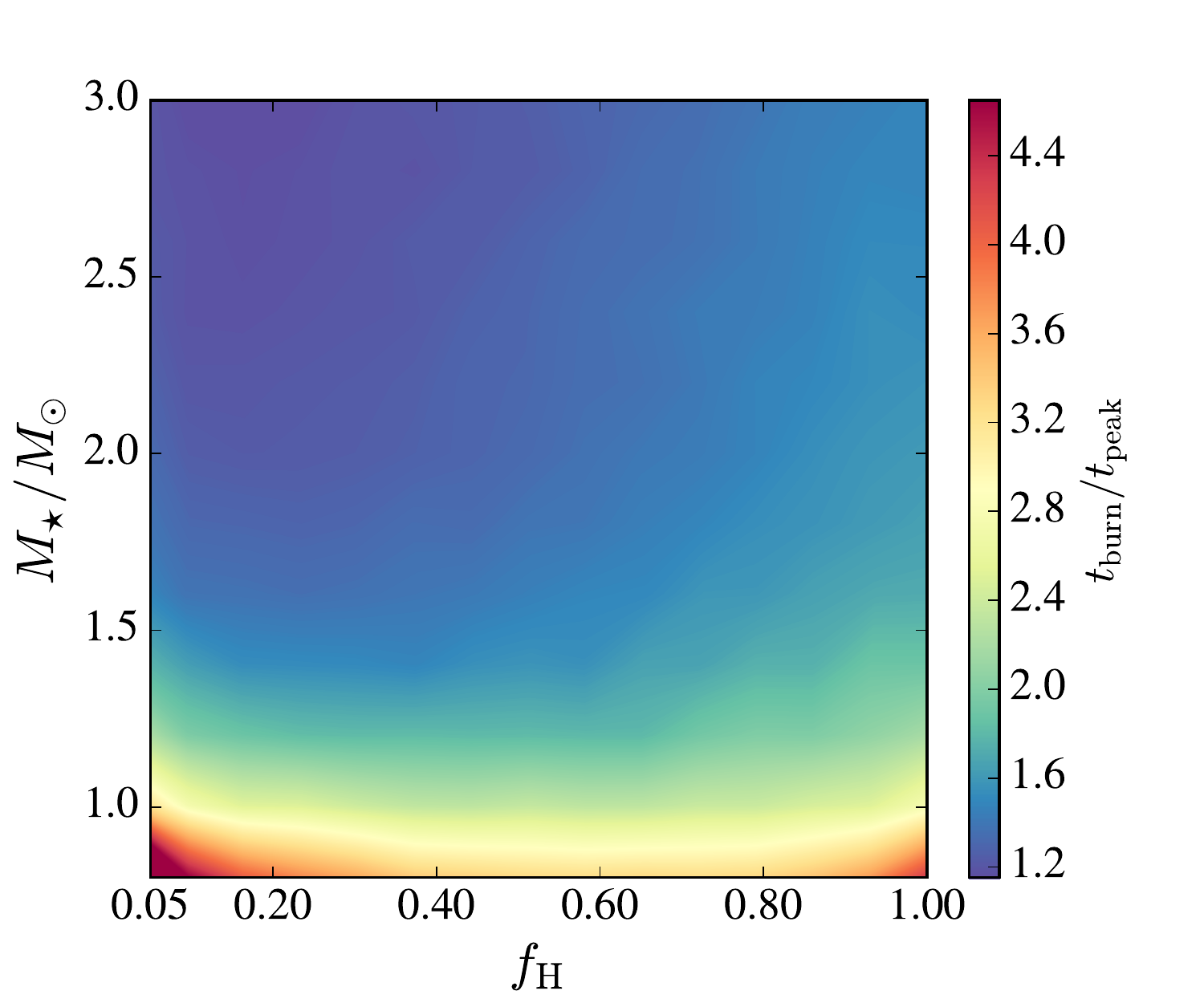}
\caption{The ratio of $t_{\rm burn}$ to $t_{\rm peak}$ as a function of $f_{\rm H}$ and stellar mass. Here $t_{\rm burn}$ is the time when 
non-solar abundance ratios begin to appear in the fallback material, specifically when  the abundance of $^{12} \rm C$ and $^{14} \rm N$  deviate from solar. We have explicitly excluded $f_{\rm H}\lesssim 0.05$ from this plot, given that these stars experience some mild contraction early in their MESA evolution. The ratio $(t_{\rm burn}/t_{\rm peak})$ reaches a maximum (minimum)  value of 7.6 (1.15) for a $0.8M_\sun$ ($3M_\sun$) star at $f_{\rm H} = 0.05$ ($f_{\rm H} = 0.23$).}
\label{fig:t_burn-t_peak}
\end{figure}

Motivated by this, in Figure~\ref{fig:t_burn-t_peak} we plot the fallback time $t_{\rm burn}$, relative to $t_{\rm peak}$, at which we expect to see anomalous abundance variations.  Here $t_{\rm burn}$ is  defined as the time at which the abundances of $^{12} \rm C$ and $^{14} \rm N$ in the fallback material, as presented in Figures \ref{fig:1Msun_xx} and \ref{fig:mdot_xx_3by3},  both deviate from unity. $t_{\rm burn} / t_{\rm peak}$ is shown in Figure~\ref{fig:t_burn-t_peak} as a function of stellar mass and age (characterized by $f_{\rm H}$). At fixed $f_{\rm H}$ we see that non-solar abundances in the fallback debris begin to appear systemically closer to $t_{\rm peak}$ as stellar mass increases. For the $3.0M_{\sun}$ star,  $t_{\rm burn} \approx 1.2 t_{\rm peak}$ for $f_{\rm H}\lesssim 0.3$. For constant $M_\star$, $t_{\rm burn} / t_{\rm peak}$ increases mildly with $f_{\rm H}$ for stars with $M_\star>$ $1.6M_{\sun}$. For stars with $M_\star<1.6M_{\sun}$, this ratio remains fairly constant  throughout the star's evolution. In summary, $t_{\rm burn}$ depends strongly  on $M_\star$ but has a relatively weak dependence on stellar age. It is important to note that independently of the mass and age of the disrupted star, no anomalous abundances are expected to be observed before $t_{\rm peak}$.

Information regarding the nature of the disrupted star should be imprinted on the properties of the TDE light curve (e.g., $t_{\rm peak}$ and $\dot{M}_{\rm peak}$) and  spectrum (particularly at $t\gtrsim t_{\rm burn}$). Current observations of TDEs show clear differences in their rise and decay properties as well as in their spectral evolution. Peculiar emission features have been observed in their spectra, which include an array of helium, hydrogen, and nitrogen  broad  line emission features. The origin of these features as well as their associated line ratios have caused significant debate.  The extreme helium to hydrogen line ratio observed in the transient event PS1-10jh was initially proposed to be the result of the tidal disruption of a helium-rich star \citep{2012Natur.485..217G}. However, such line ratios have been shown to  arise  naturally from  the reprocessing of radiation through the  fallback debris of a disrupted Sun-like star \citep{2016ApJ...827....3R}.  As for the additional presence of rare nitrogen features, \citet{2016MNRAS.458..127K} first proposed that the disruption of MS stars with evolved stellar compositions could lead to enhanced nitrogen (as well as anomalous helium and carbon abundances). 

\begin{figure}[tbp]
\epsscale{1.23}
\plotone{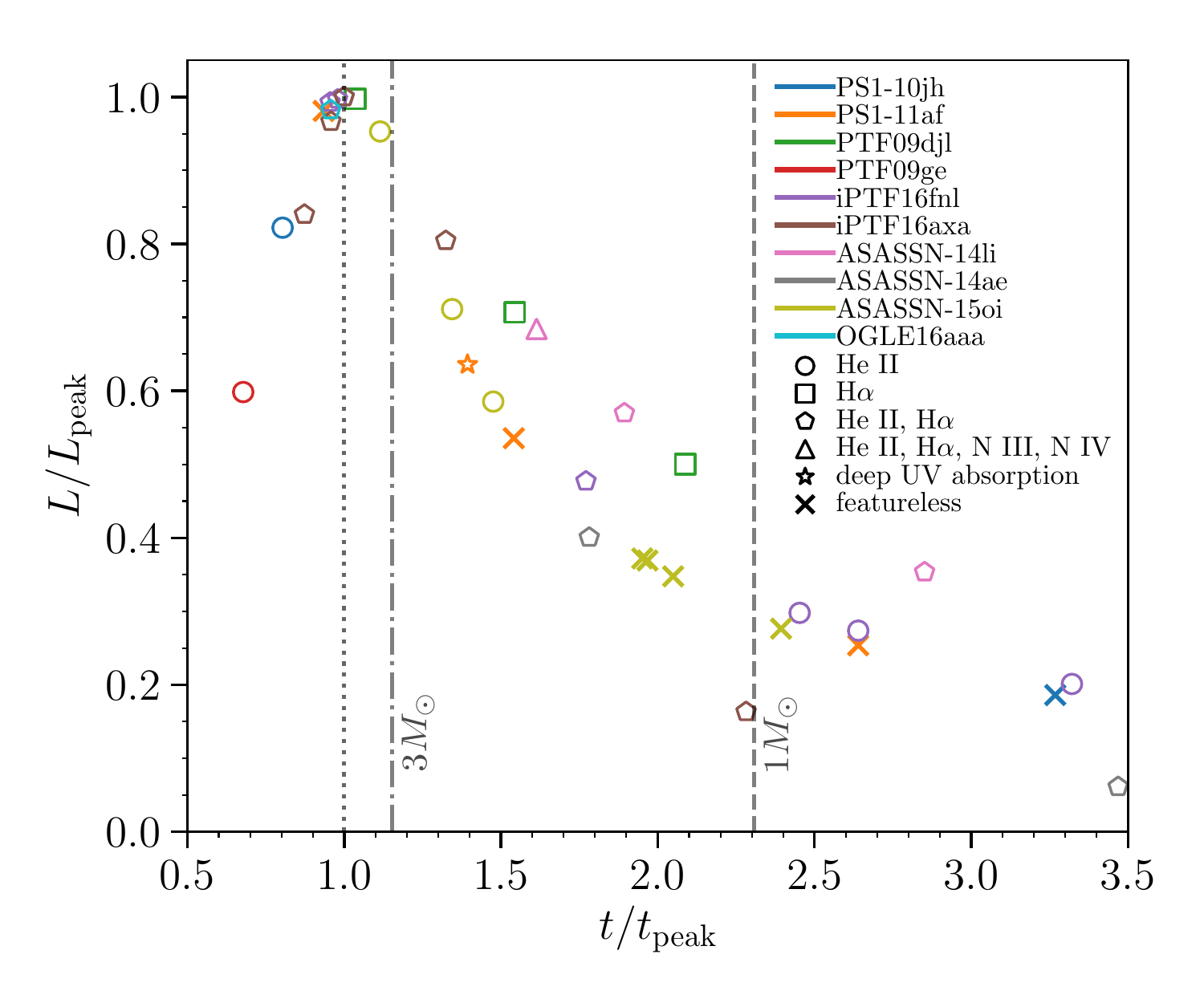}
\caption{Compositional features in the spectra of well-sampled tidal disruption events with existing spectroscopic observations. The $y$- and $x$-axes show luminosity and time relative to peak respectively,  with different colors corresponding to distinct events, and different symbols corresponding to different spectral features. We show the minimum values of $t_{\rm burn}/t_{\rm peak}$ (Figure \ref{fig:t_burn-t_peak}) as derived from our study for a $1M_\sun$ ({\it dashed} line) and $3M_\sun$ ({\it dash-dotted} line) star.
}
\label{fig:spectra}
\end{figure}

In Figure~\ref{fig:spectra}, we show compositional features in the spectra of ten observed TDEs. We place each spectrum in the light curve of each event, relative to its peak luminosity and peak time. Symbols indicate features present in the spectra. Bolometric light curve fits for each event are from \citet{2018Mockler}. Data is taken from \citet{2012Natur.485..217G,2015ApJ...815L...5G,2014ApJ...780...44C,2014ApJ...793...38A,2018MNRAS.473.1130B,2017ApJ...844...46B,2017ApJ...842...29H,2014MNRAS.445.3263H,2016MNRAS.463.3813H,2016MNRAS.455.2918H,2016ApJ...818L..32C,2016MNRAS.462.3993B,2017MNRAS.466.4904B,2016NatAs...1E..34L,2017MNRAS.465L.114W}.
Note that this figure shows TDEs with well-sampled light curves and existing spectroscopic observations. Several TDE spectra show compositional features at or near the peak in their light curve. Our calculations (in particular see Section \ref{sec: evolved ms star disruption} and Figure \ref{fig:t_burn-t_peak}) predict no compositional abundance changes (relative to solar) in the fallback material at or near 
peak due to the star.  This implies that the strong suppression of hydrogen Balmer line emission relative to helium line emission should occur even at solar composition, as argued by \citet{2016ApJ...827....3R}, due to optical depth effects alone. For observations at $t> t_{\rm burn}$, we expect the reprocessing material to be enhanced in helium, yet the optical depth effects  are expected to be less important \citep{2014ApJ...783...23G}. As such,  radiation transfer calculations are needed before firm conclusions can be derived from observations of evolving line ratios in a given TDE.

Nitrogen emission lines, on the other hand, are only currently detected at $\approx 1.5 t_{\rm peak}$. If their presence is primarily  attributed to a drastic increase in nitrogen abundance, then based on the results shown in  Figures~\ref{fig:LT} and  \ref{fig:t_burn-t_peak}, one would conclude that $M_\star \gtrsim 1.8M_\odot$ for the star whose disruption triggered the ASASSN-14li flaring event. This timescale for chemical enrichment (i.e., $t_{\rm burn}$)  can thus provide a direct observational test of which stars are being disrupted by the central SMBH. 

Much progress has been made in understanding how the feeding rate onto a SMBH proceeds after the disruption of a particular star, and in deriving the generic properties of the flares that follow from this. There still remain a number of mysteries, especially concerning the identity of the star, the nature of the energy dissipation mechanism, and the time scales involved. The modeling of the flare itself (i.e., the dissipation  mechanism and the radiation processes) is a formidable challenge to theorists and to numerical techniques. It is also a challenge for observers, in their quest to detect fine details in  distant, fading  sources. The class of models we have presented here predict that the spectral properties of the fading signals will turn out to be even more telling and fascinating that initially anticipated.

Future work will include a more detailed exploration of the parameters governing the abundance of the fallback material, including hydrodynamical calculations \citep[e.g.,][]{2017ApJ...841..132L} as well as radiative transfer calculations \citep[e.g.,][]{2016ApJ...827....3R} evolved over time for different properties of the reprocessing material. Studies of this sort, in comparison with improved spectral observations of TDEs, will undoubtedly help clarify the physics governing these transient sources.

\acknowledgements
We thank B. Mockler for insightful conversations, as well as for providing bolometric light curve fits for the TDEs shown in Figure~\ref{fig:spectra}.
We thank N. Roth, J. Guillochon, R. Foley and D. Kasen for useful discussions. M. G.-G. and E.R.-R. are grateful for support from the Packard Foundation and from Julie Packard. J. L.-S. and E. R.-R. acknowledge support from NASA ATP grant NNX14AH37G and NSF grant AST-1615881.

\bibliography{refs}

\begin{thebibliography}{}
\expandafter\ifx\csname natexlab\endcsname\relax\def\natexlab#1{#1}\fi
\providecommand{\url}[1]{\href{#1}{#1}}

\bibitem[{{Arcavi} {et~al.}(2014){Arcavi}, {Gal-Yam}, {Sullivan}, {Pan},
  {Cenko}, {Horesh}, {Ofek}, {De Cia}, {Yan}, {Yang}, {Howell}, {Tal},
  {Kulkarni}, {Tendulkar}, {Tang}, {Xu}, {Sternberg}, {Cohen}, {Bloom},
  {Nugent}, {Kasliwal}, {Perley}, {Quimby}, {Miller}, {Theissen}, \&
  {Laher}}]{2014ApJ...793...38A}
{Arcavi}, I., {Gal-Yam}, A., {Sullivan}, M., {et~al.} 2014, \apj, 793, 38

\bibitem[{{Asplund} {et~al.}(2009){Asplund}, {Grevesse}, {Sauval}, \&
  {Scott}}]{2009ARA&A..47..481A}
{Asplund}, M., {Grevesse}, N., {Sauval}, A.~J., \& {Scott}, P. 2009, \araa, 47,
  481

\bibitem[{{Auchettl} {et~al.}(2017){Auchettl}, {Guillochon}, \&
  {Ramirez-Ruiz}}]{2017ApJ...838..149A}
{Auchettl}, K., {Guillochon}, J., \& {Ramirez-Ruiz}, E. 2017, \apj, 838, 149

\bibitem[{{Blagorodnova} {et~al.}(2017){Blagorodnova}, {Gezari}, {Hung},
  {Kulkarni}, {Cenko}, {Pasham}, {Yan}, {Arcavi}, {Ben-Ami}, {Bue}, {Cantwell},
  {Cao}, {Castro-Tirado}, {Fender}, {Fremling}, {Gal-Yam}, {Ho}, {Horesh},
  {Hosseinzadeh}, {Kasliwal}, {Kong}, {Laher}, {Leloudas}, {Lunnan}, {Masci},
  {Mooley}, {Neill}, {Nugent}, {Powell}, {Valeev}, {Vreeswijk}, {Walters}, \&
  {Wozniak}}]{2017ApJ...844...46B}
{Blagorodnova}, N., {Gezari}, S., {Hung}, T., {et~al.} 2017, \apj, 844, 46

\bibitem[{{Bonnerot} {et~al.}(2016){Bonnerot}, {Rossi}, {Lodato}, \&
  {Price}}]{2016MNRAS.455.2253B}
{Bonnerot}, C., {Rossi}, E.~M., {Lodato}, G., \& {Price}, D.~J. 2016, \mnras,
  455, 2253

\bibitem[{{Brown} {et~al.}(2015){Brown}, {Levan}, {Stanway}, {Tanvir}, {Cenko},
  {Berger}, {Chornock}, \& {Cucchiaria}}]{2015MNRAS.452.4297B}
{Brown}, G.~C., {Levan}, A.~J., {Stanway}, E.~R., {et~al.} 2015, \mnras, 452,
  4297

\bibitem[{{Brown} {et~al.}(2017){Brown}, {Holoien}, {Auchettl}, {Stanek},
  {Kochanek}, {Shappee}, {Prieto}, \& {Grupe}}]{2017MNRAS.466.4904B}
{Brown}, J.~S., {Holoien}, T.~W.-S., {Auchettl}, K., {et~al.} 2017, \mnras,
  466, 4904

\bibitem[{{Brown} {et~al.}(2016){Brown}, {Shappee}, {Holoien}, {Stanek},
  {Kochanek}, \& {Prieto}}]{2016MNRAS.462.3993B}
{Brown}, J.~S., {Shappee}, B.~J., {Holoien}, T.~W.-S., {et~al.} 2016, \mnras,
  462, 3993

\bibitem[{{Brown} {et~al.}(2018){Brown}, {Kochanek}, {Holoien}, {Stanek},
  {Auchettl}, {Shappee}, {Prieto}, {Morrell}, {Falco}, {Strader}, {Chomiuk},
  {Post}, {Villanueva}, {Mathur}, {Dong}, {Chen}, \&
  {Bose}}]{2018MNRAS.473.1130B}
{Brown}, J.~S., {Kochanek}, C.~S., {Holoien}, T.~W.-S., {et~al.} 2018, \mnras,
  473, 1130

\bibitem[{{Cenko} {et~al.}(2012){Cenko}, {Krimm}, {Horesh}, {Rau}, {Frail},
  {Kennea}, {Levan}, {Holland}, {Butler}, {Quimby}, {Bloom}, {Filippenko},
  {Gal-Yam}, {Greiner}, {Kulkarni}, {Ofek}, {Olivares E.}, {Schady},
  {Silverman}, {Tanvir}, \& {Xu}}]{2012ApJ...753...77C}
{Cenko}, S.~B., {Krimm}, H.~A., {Horesh}, A., {et~al.} 2012, \apj, 753, 77

\bibitem[{{Cenko} {et~al.}(2016){Cenko}, {Cucchiara}, {Roth}, {Veilleux},
  {Prochaska}, {Yan}, {Guillochon}, {Maksym}, {Arcavi}, {Butler}, {Filippenko},
  {Fruchter}, {Gezari}, {Kasen}, {Levan}, {Miller}, {Pasham}, {Ramirez-Ruiz},
  {Strubbe}, {Tanvir}, \& {Tombesi}}]{2016ApJ...818L..32C}
{Cenko}, S.~B., {Cucchiara}, A., {Roth}, N., {et~al.} 2016, \apjl, 818, L32

\bibitem[{{Chornock} {et~al.}(2014){Chornock}, {Berger}, {Gezari}, {Zauderer},
  {Rest}, {Chomiuk}, {Kamble}, {Soderberg}, {Czekala}, {Dittmann}, {Drout},
  {Foley}, {Fong}, {Huber}, {Kirshner}, {Lawrence}, {Lunnan}, {Marion},
  {Narayan}, {Riess}, {Roth}, {Sanders}, {Scolnic}, {Smartt}, {Smith},
  {Stubbs}, {Tonry}, {Burgett}, {Chambers}, {Flewelling}, {Hodapp}, {Kaiser},
  {Magnier}, {Martin}, {Neill}, {Price}, \& {Wainscoat}}]{2014ApJ...780...44C}
{Chornock}, R., {Berger}, E., {Gezari}, S., {et~al.} 2014, \apj, 780, 44

\bibitem[{{Evans} \& {Kochanek}(1989)}]{1989ApJ...346L..13E}
{Evans}, C.~R., \& {Kochanek}, C.~S. 1989, \apjl, 346, L13

\bibitem[{{Frank} \& {Rees}(1976)}]{1976MNRAS.176..633F}
{Frank}, J., \& {Rees}, M.~J. 1976, \mnras, 176, 633

\bibitem[{{French} {et~al.}(2016){French}, {Arcavi}, \&
  {Zabludoff}}]{2016ApJ...818L..21F}
{French}, K.~D., {Arcavi}, I., \& {Zabludoff}, A. 2016, \apjl, 818, L21

\bibitem[{{French} {et~al.}(2017){French}, {Arcavi}, \&
  {Zabludoff}}]{2017ApJ...835..176F}
---. 2017, \apj, 835, 176

\bibitem[{{Gezari} {et~al.}(2015){Gezari}, {Chornock}, {Lawrence}, {Rest},
  {Jones}, {Berger}, {Challis}, \& {Narayan}}]{2015ApJ...815L...5G}
{Gezari}, S., {Chornock}, R., {Lawrence}, A., {et~al.} 2015, \apjl, 815, L5

\bibitem[{{Gezari} {et~al.}(2012){Gezari}, {Chornock}, {Rest}, {Huber},
  {Forster}, {Berger}, {Challis}, {Neill}, {Martin}, {Heckman}, {Lawrence},
  {Norman}, {Narayan}, {Foley}, {Marion}, {Scolnic}, {Chomiuk}, {Soderberg},
  {Smith}, {Kirshner}, {Riess}, {Smartt}, {Stubbs}, {Tonry}, {Wood-Vasey},
  {Burgett}, {Chambers}, {Grav}, {Heasley}, {Kaiser}, {Kudritzki}, {Magnier},
  {Morgan}, \& {Price}}]{2012Natur.485..217G}
{Gezari}, S., {Chornock}, R., {Rest}, A., {et~al.} 2012, \nat, 485, 217

\bibitem[{{Guillochon} {et~al.}(2014){Guillochon}, {Manukian}, \&
  {Ramirez-Ruiz}}]{2014ApJ...783...23G}
{Guillochon}, J., {Manukian}, H., \& {Ramirez-Ruiz}, E. 2014, \apj, 783, 23

\bibitem[{{Guillochon} \& {Ramirez-Ruiz}(2013)}]{2013ApJ...767...25G}
{Guillochon}, J., \& {Ramirez-Ruiz}, E. 2013, \apj, 767, 25

\bibitem[{{Guillochon} \& {Ramirez-Ruiz}(2015)}]{2015ApJ...809..166G}
---. 2015, \apj, 809, 166

\bibitem[{{Guillochon} {et~al.}(2009){Guillochon}, {Ramirez-Ruiz}, {Rosswog},
  \& {Kasen}}]{2009ApJ...705..844G}
{Guillochon}, J., {Ramirez-Ruiz}, E., {Rosswog}, S., \& {Kasen}, D. 2009, \apj,
  705, 844

\bibitem[{{Hayasaki} {et~al.}(2016){Hayasaki}, {Stone}, \&
  {Loeb}}]{2016MNRAS.461.3760H}
{Hayasaki}, K., {Stone}, N., \& {Loeb}, A. 2016, \mnras, 461, 3760

\bibitem[{{Holoien} {et~al.}(2014){Holoien}, {Prieto}, {Bersier}, {Kochanek},
  {Stanek}, {Shappee}, {Grupe}, {Basu}, {Beacom}, {Brimacombe}, {Brown},
  {Davis}, {Jencson}, {Pojmanski}, \& {Szczygie{\l}}}]{2014MNRAS.445.3263H}
{Holoien}, T.~W.-S., {Prieto}, J.~L., {Bersier}, D., {et~al.} 2014, \mnras,
  445, 3263

\bibitem[{{Holoien} {et~al.}(2016{\natexlab{a}}){Holoien}, {Kochanek},
  {Prieto}, {Grupe}, {Chen}, {Godoy-Rivera}, {Stanek}, {Shappee}, {Dong},
  {Brown}, {Basu}, {Beacom}, {Bersier}, {Brimacombe}, {Carlson}, {Falco},
  {Johnston}, {Madore}, {Pojmanski}, \& {Seibert}}]{2016MNRAS.463.3813H}
{Holoien}, T.~W.-S., {Kochanek}, C.~S., {Prieto}, J.~L., {et~al.}
  2016{\natexlab{a}}, \mnras, 463, 3813

\bibitem[{{Holoien} {et~al.}(2016{\natexlab{b}}){Holoien}, {Kochanek},
  {Prieto}, {Stanek}, {Dong}, {Shappee}, {Grupe}, {Brown}, {Basu}, {Beacom},
  {Bersier}, {Brimacombe}, {Danilet}, {Falco}, {Guo}, {Jose}, {Herczeg},
  {Long}, {Pojmanski}, {Simonian}, {Szczygie{\l}}, {Thompson}, {Thorstensen},
  {Wagner}, \& {Wo{\'z}niak}}]{2016MNRAS.455.2918H}
---. 2016{\natexlab{b}}, \mnras, 455, 2918

\bibitem[{{Hung} {et~al.}(2017){Hung}, {Gezari}, {Blagorodnova}, {Roth},
  {Cenko}, {Kulkarni}, {Horesh}, {Arcavi}, {McCully}, {Yan}, {Lunnan},
  {Fremling}, {Cao}, {Nugent}, \& {Wozniak}}]{2017ApJ...842...29H}
{Hung}, T., {Gezari}, S., {Blagorodnova}, N., {et~al.} 2017, \apj, 842, 29

\bibitem[{{Kesden}(2012)}]{2012PhRvD..86f4026K}
{Kesden}, M. 2012, \prd, 86, 064026

\bibitem[{{Kippenhahn} {et~al.}(2012){Kippenhahn}, {Weigert}, \&
  {Weiss}}]{2012sse..book.....K}
{Kippenhahn}, R., {Weigert}, A., \& {Weiss}, A. 2012, {Stellar Structure and
  Evolution}, doi:10.1007/978-3-642-30304-3

\bibitem[{{Kochanek}(2016)}]{2016MNRAS.458..127K}
{Kochanek}, C.~S. 2016, \mnras, 458, 127

\bibitem[{{Komossa}(2015)}]{2015JHEAp...7..148K}
{Komossa}, S. 2015, Journal of High Energy Astrophysics, 7, 148

\bibitem[{{Laguna} {et~al.}(1993){Laguna}, {Miller}, {Zurek}, \&
  {Davies}}]{1993ApJ...410L..83L}
{Laguna}, P., {Miller}, W.~A., {Zurek}, W.~H., \& {Davies}, M.~B. 1993, \apjl,
  410, L83

\bibitem[{{Law-Smith} {et~al.}(2017{\natexlab{a}}){Law-Smith}, {MacLeod},
  {Guillochon}, {Macias}, \& {Ramirez-Ruiz}}]{2017ApJ...841..132L}
{Law-Smith}, J., {MacLeod}, M., {Guillochon}, J., {Macias}, P., \&
  {Ramirez-Ruiz}, E. 2017{\natexlab{a}}, \apj, 841, 132

\bibitem[{{Law-Smith} {et~al.}(2017{\natexlab{b}}){Law-Smith}, {Ramirez-Ruiz},
  {Ellison}, \& {Foley}}]{2017ApJ...850...22L}
{Law-Smith}, J., {Ramirez-Ruiz}, E., {Ellison}, S.~L., \& {Foley}, R.~J.
  2017{\natexlab{b}}, \apj, 850, 22

\bibitem[{{Leloudas} {et~al.}(2016){Leloudas}, {Fraser}, {Stone}, {van Velzen},
  {Jonker}, {Arcavi}, {Fremling}, {Maund}, {Smartt}, {Kr{\`i}hler},
  {Miller-Jones}, {Vreeswijk}, {Gal-Yam}, {Mazzali}, {De Cia}, {Howell},
  {Inserra}, {Patat}, {de Ugarte Postigo}, {Yaron}, {Ashall}, {Bar},
  {Campbell}, {Chen}, {Childress}, {Elias-Rosa}, {Harmanen}, {Hosseinzadeh},
  {Johansson}, {Kangas}, {Kankare}, {Kim}, {Kuncarayakti}, {Lyman}, {Magee},
  {Maguire}, {Malesani}, {Mattila}, {McCully}, {Nicholl}, {Prentice},
  {Romero-Ca{\~n}izales}, {Schulze}, {Smith}, {Sollerman}, {Sullivan},
  {Tucker}, {Valenti}, {Wheeler}, \& {Young}}]{2016NatAs...1E..34L}
{Leloudas}, G., {Fraser}, M., {Stone}, N.~C., {et~al.} 2016, Nature Astronomy,
  1, 0034

\bibitem[{{Lodato} {et~al.}(2009){Lodato}, {King}, \&
  {Pringle}}]{2009MNRAS.392..332L}
{Lodato}, G., {King}, A.~R., \& {Pringle}, J.~E. 2009, \mnras, 392, 332

\bibitem[{{MacLeod} {et~al.}(2012){MacLeod}, {Guillochon}, \&
  {Ramirez-Ruiz}}]{2012ApJ...757..134M}
{MacLeod}, M., {Guillochon}, J., \& {Ramirez-Ruiz}, E. 2012, \apj, 757, 134

\bibitem[{{Magorrian} \& {Tremaine}(1999)}]{1999MNRAS.309..447M}
{Magorrian}, J., \& {Tremaine}, S. 1999, \mnras, 309, 447

\bibitem[{{Merloni} {et~al.}(2015){Merloni}, {Dwelly}, {Salvato},
  {Georgakakis}, {Greiner}, {Krumpe}, {Nandra}, {Ponti}, \&
  {Rau}}]{2015MNRAS.452...69M}
{Merloni}, A., {Dwelly}, T., {Salvato}, M., {et~al.} 2015, \mnras, 452, 69

\bibitem[{{Mockler} {et~al.}(2018){Mockler}, {Guillochon}, \&
  {Ramirez-Ruiz}}]{2018Mockler}
{Mockler}, B., {Guillochon}, J., \& {Ramirez-Ruiz}, E. 2018, submitted to ApJ

\bibitem[{{Paxton} {et~al.}(2011){Paxton}, {Bildsten}, {Dotter}, {Herwig},
  {Lesaffre}, \& {Timmes}}]{2011ApJS..192....3P}
{Paxton}, B., {Bildsten}, L., {Dotter}, A., {et~al.} 2011, \apjs, 192, 3

\bibitem[{{Phinney}(1989)}]{1989IAUS..136..543P}
{Phinney}, E.~S. 1989, in IAU Symposium, Vol. 136, The Center of the Galaxy,
  ed. M.~{Morris}, 543

\bibitem[{{Ramirez-Ruiz} \& {Rosswog}(2009)}]{2009ApJ...697L..77R}
{Ramirez-Ruiz}, E., \& {Rosswog}, S. 2009, \apjl, 697, L77

\bibitem[{{Rees}(1988)}]{1988Natur.333..523R}
{Rees}, M.~J. 1988, \nat, 333, 523

\bibitem[{{Roth} {et~al.}(2016){Roth}, {Kasen}, {Guillochon}, \&
  {Ramirez-Ruiz}}]{2016ApJ...827....3R}
{Roth}, N., {Kasen}, D., {Guillochon}, J., \& {Ramirez-Ruiz}, E. 2016, \apj,
  827, 3

\bibitem[{{Saxton} {et~al.}(2012){Saxton}, {Read}, {Esquej}, {Komossa},
  {Dougherty}, {Rodriguez-Pascual}, \& {Barrado}}]{2012A&A...541A.106S}
{Saxton}, R.~D., {Read}, A.~M., {Esquej}, P., {et~al.} 2012, \aap, 541, A106

\bibitem[{{Shiokawa} {et~al.}(2015){Shiokawa}, {Krolik}, {Cheng}, {Piran}, \&
  {Noble}}]{2015ApJ...804...85S}
{Shiokawa}, H., {Krolik}, J.~H., {Cheng}, R.~M., {Piran}, T., \& {Noble}, S.~C.
  2015, \apj, 804, 85

\bibitem[{{Strubbe} \& {Quataert}(2009)}]{2009MNRAS.400.2070S}
{Strubbe}, L.~E., \& {Quataert}, E. 2009, \mnras, 400, 2070

\bibitem[{{Wyrzykowski} {et~al.}(2017){Wyrzykowski}, {Zieli{\'n}ski},
  {Kostrzewa-Rutkowska}, {Hamanowicz}, {Jonker}, {Arcavi}, {Guillochon},
  {Brown}, {Koz{\l}owski}, {Udalski}, {Szyma{\'n}ski}, {Soszy{\'n}ski},
  {Poleski}, {Pietrukowicz}, {Skowron}, {Mr{\'o}z}, {Ulaczyk}, {Pawlak},
  {Rybicki}, {Greiner}, {Kr{\"u}hler}, {Bolmer}, {Smartt}, {Maguire}, \&
  {Smith}}]{2017MNRAS.465L.114W}
{Wyrzykowski}, {\L}., {Zieli{\'n}ski}, M., {Kostrzewa-Rutkowska}, Z., {et~al.}
  2017, \mnras, 465, L114

\end{thebibliography}
\bibliographystyle{aasjournal}

\end{document}